\documentclass[11pt,oneside]{article}

\usepackage{palatino}
\usepackage[margin=1in]{geometry}
\usepackage{color,xcolor,soul}
\usepackage[abspage,user,savepos]{zref}
\usepackage[colorlinks, citecolor=blue]{hyperref}  
\usepackage{natbib}
\usepackage{fp}
\usepackage{framed}
\usepackage{cuted,balance}
\usepackage{setspace}
\usepackage{subcaption}

\usepackage{graphicx} 
\usepackage{epsfig} 
\usepackage{mathtools}
\usepackage{dsfont,amsmath} 
\usepackage{amssymb}  
\usepackage{algorithm,algorithmic}
\usepackage[capitalize]{cleveref}

\DeclareMathOperator*{\argmax}{argmax}
\newcommand{\Exp}{\mathrm{E}}
\newcommand{\sBR}{\underline{\mathrm{br}}}
\newcommand{\BRe}{\mathrm{br}_{\epsilon}}
\newcommand{\BR}{\mathrm{br}}
\newcommand{\val}{\underline{\mathrm{val}}}
\newcommand{\maxmin}{\mathrm{val}}
\newcommand{\be}{\begin{equation}}
\newcommand{\ee}{\end{equation}}
\newcommand{\nn}{\nonumber}

\newcommand{\uw}{\underline{w}}
\newcommand{\ow}{\overline{w}}
\newcommand{\uomega}{\underline{\omega}}
\newcommand{\oomega}{\overline{\omega}}

\newcommand{\ue}{\underline{e}}

\newcommand{\bdelta}{\bar{\delta}}
\newcommand{\bc}{\bar{c}}
\newcommand{\bQ}{\bar{Q}}
\newcommand{\bmu}{\bar{\mu}}
\newcommand{\tv}{\tilde{v}}
\newcommand{\ane}{\mathrm{a.n.e.}}

\makeatletter
  \renewcommand{\ALG@name}{Algorithm Family}
  \makeatother

\newcounter{lemma}
\newenvironment{lemma}{\refstepcounter{lemma}\par\medskip
\noindent 
\textbf{Lemma \thelemma.} \em \rmfamily}
{\medskip}

\newcounter{proposition}
{\medskip}

\newenvironment{proposition*}[1]{\refstepcounter{proposition}\par\medskip
\noindent 
\textbf{Proposition \theproposition~(#1)} \em \rmfamily}
{\medskip}

\newcounter{theorem}
\newenvironment{theorem}{\refstepcounter{theorem}\par\medskip
\noindent 
\textbf{Theorem \thetheorem.} \em \rmfamily}
{\medskip}

\newcounter{corollary}
\newenvironment{corollary}{\refstepcounter{corollary}\par\medskip
\noindent 
\textbf{Corollary \thecorollary.} \em \rmfamily}
{\medskip}

\newcounter{definition}
\newenvironment{definition}[1]{\refstepcounter{definition}\par\medskip
\noindent 
\textbf{Definition \thedefinition~(#1)} \em \rmfamily}
{\medskip}
\newenvironment{definition*}{\refstepcounter{definition}\par\medskip
\noindent 
\textbf{Definition \thedefinition.} \em \rmfamily}
{\medskip}

\newcounter{assumption}
\newenvironment{assumption}{\refstepcounter{assumption}\par\medskip
\noindent 
\textbf{Assumption \theassumption.} \em \rmfamily}
{\medskip}

\newcounter{remark}
\newenvironment{remark}{\refstepcounter{remark}\par\medskip
\noindent 
\textbf{Remark \theremark.} \em \rmfamily}{}
{\medskip}

\newcounter{example}
{\medskip}

\newenvironment{myproof}{
   \indent \textit{Proof.} \rmfamily}{\hfill $\square$}
   
\begin{document}

\date{}

\title{\LARGE \bf
Convergence of Heterogeneous Learning Dynamics in Zero-sum Stochastic Games
}

\author{Yuksel Arslantas, Ege Yuceel, Yigit Yalin, and Muhammed O. Sayin
\thanks{Y. Arslantas, E. Yuceel, and M. O. Sayin are with the Department of Electrical \& Electronics Engineering, and Y. Yalin is with the Department of Computer Engineering at Bilkent University, Ankara, T\"{u}rkiye 06800. (Emails: {\tt\small yuksel.arslantas@bilkent.edu.tr},  {\tt\small ege.yuceel@ug.bilkent.edu.tr},  {\tt\small yigit.yalin@ug.bilkent.edu.tr},  {\tt\small sayin@ee.bilkent.edu.tr})}%
}
\maketitle

\bigskip

\begin{center}
\textbf{Abstract}
\end{center}
This paper presents new families of algorithms for the repeated play of two-agent (near) zero-sum games and two-agent zero-sum stochastic games. For example, the family includes fictitious play and its variants as members. Commonly, the algorithms in this family are all uncoupled, rational, and convergent even in heterogeneous cases, e.g., where the dynamics may differ in terms of learning rates, full, none or temporal access to opponent actions, and model-based vs model-free learning. The convergence of heterogeneous dynamics is of practical interest especially in competitive environments since agents may have no means or interests in following the same dynamic with the same parameters. We prove that any mixture of such asymmetries does not impact the algorithms' convergence to equilibrium (or near equilibrium if there is experimentation) in zero-sum games with repeated play and in zero-sum (irreducible) stochastic games with sufficiently small discount factors.

\begin{spacing}{1.245}

\section{Introduction}
\label{sec:intro}

Recently, provable convergence of multi-agent reinforcement learning in stochastic games (also known as Markov games) has attracted attention due to their success in wide range of applications for artificial intelligence, e.g., see \citep{ref:Arslan17,ref:Daskalakis20,ref:Leslie20,ref:Sayin20,ref:Sayin21,ref:Sayin22,ref:Wei21,ref:Baudin22,ref:Baudin22b}. This line of work focuses on (radically) uncoupled, rational and convergent learning. However, the existing convergence results for zero-sum games focus mainly on \textit{homogenous} settings where every agent follow the same learning dynamic. The homogeneity assumption limits the predictive power of these findings to address multi-agent interactions in practice because non-cooperative agents may not have any interest or means to follow the same dynamic, especially in competitive environments. There is a critical need for a paradigm shift toward addressing the provable convergence in \textit{heterogeneous} settings where agents may follow different learning dynamics.  

\begin{figure}[t!]
  \centering
  \includegraphics[width=.65\textwidth]{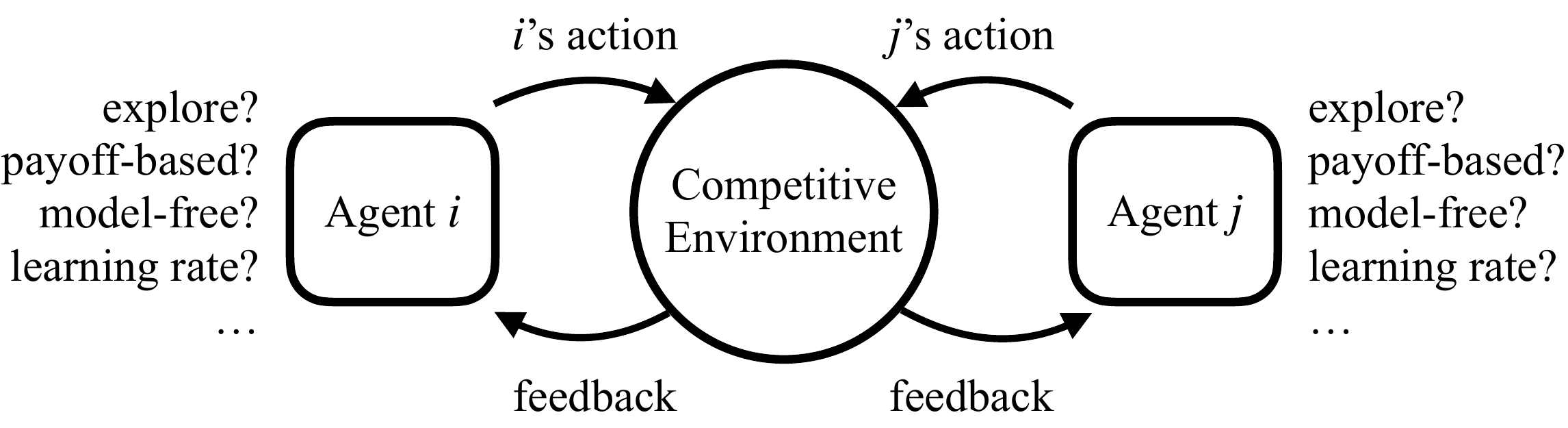}
  \caption{An illustration highlighting the differences in the interests and means of agents interacting with each other in a competitive environment. }\label{fig:motivation}
  \end{figure}

\textbf{Motivating Examples.} 
Consider two agents interacting with each other in a competitive environment, as illustrated in Fig.\ref{fig:motivation}. The agents are taking actions to maximize their (misaligned) payoffs while their joint actions determine the payoffs they receive. In such cases, finding the best action is not a well-defined optimization problem for agents to solve through introspective reasoning due to its dependence on the opponent's play. Playing the worst-case equilibrium strategy can be very conservative against opponents with limited cognitive and computational capabilities. Instead, agents can learn and adapt their play by interacting with each other while receiving feedback from the environment \citep{ref:Fudenberg09}. However, how they play depends on their interests and means in terms of toleration to exploration, information structure, and model knowledge, e.g., see Fig. \ref{fig:motivation}. Any differences in their interests and means inevitably lead to heterogeneous dynamics.

\textbf{Challenges.} 
One of the challenges is to establish a unifying framework to address such heterogeneities and the challenge gets elevated for stochastic games (SGs) compared to the repeated play of games. For example, we can view stochastic games as agents are playing \textit{stage games} specific to each state whenever the associated state gets visited. However, the payoffs of these stage games depend also on how the agents would play in future stages as the actions taken determine not only the rewards received in the current stage but also the probability of transition to the next state, and therefore, the \textit{continuation payoff}. Recently, two-timescale learning framework has addressed the possible non-stationarity of the stage-game payoffs \citep{ref:Leslie20,ref:Sayin20,ref:Sayin21}. However, these stage game payoffs may not sum to zero depending on how agents estimate the continuation payoff. The deviation of the stage games from the zero-sum structure gets boosted under heterogeneities. Hence, there is a need for sharp convergence guarantees for heterogeneous dynamics also for near zero-sum games. 

\textbf{Contributions.} We present new families of algorithms for the repeated play of two-agent zero-sum games and two-agent zero-sum stochastic games. The families provide a unifying framework to address heterogeneities in terms of exploration, access to opponent actions, model knowledge and step sizes. We focus on best-response-type learning dynamics, where agents respond greedily to the belief they formed about how the opponent plays under the assumption that the opponent plays according to some stationary strategy. Full and no access to opponent actions, resp., correspond to belief-based and payoff-based dynamics, while temporal access is the cases where the agent observes the opponent action with some probability independent of the history of the game. Belief-based and payoff-based dynamics have been studied extensively in the multi-agent learning literature \citep{ref:Fudenberg09,ref:Busoniu10}. However, we are not aware of any previous work addressing the temporal access though loss of messages or line-of-sight are common issues in communication/control applications, e.g., see \citep{ref:Yuksel13}. 

The family for the repeated play of the games includes fictitious play \citep{ref:Brown51}, smoothed fictitious play \citep{ref:Fudenberg93}, and individual Q-learning \citep{ref:Leslie05}. We also extend the family to stochastic games based on the \textit{stage-game framework} where agents play stage games as if their payoffs are the Q-functions they estimate. The family for stochastic games includes the new variants of fictitious play, introduced by \citep{ref:Sayin20,ref:Sayin21}. The algorithm families presented also have the following desired properties for multi-agent learning:
\begin{itemize}
\item \textit{(Radically) uncoupled:} The dynamics are game-agnostic by not requiring the knowledge of the opponent's objective. The dynamics also may not require the knowledge of the opponent's actions and the underlying model.
\item \textit{Rational:} The dynamics can achieve the (near) best performance against opponents following stationary strategies under standard assumptions on the step sizes (and the transition kernel for SGs), see Corollaries \ref{cor:rationalG} and \ref{cor:rationalSG}.
\item \textit{Convergent in homogeneous and heterogeneous cases:} Any mixture of the dynamics from the algorithm families reaches (near) equilibrium almost surely in zero-sum (stochastic) games under the same set of assumptions and for sufficiently small discount factor if the agents use different step sizes, see Corollary \ref{cor:G} and Theorem \ref{thm:SG}.
\end{itemize}
Notably, we can approximate the discrete-time update of fictitious play and its variants with continuous-time best-response dynamics and characterize the convergence properties of the former by formulating Lyapunov function for the latter based on stochastic approximation methods, e.g., see \citep{ref:Benaim05,ref:Perkins13}. To this end, we formulate a unifying Lyapunov function addressing heterogeneities even for near zero-sum games. 

\textbf{Related Works.} 
Whether non-equilibrium adaptation of learning agents would reach equilibrium has been studied extensively in the learning-in-games literature \citep{ref:Fudenberg09}. For example, fictitious play \citep{ref:Brown51} and its variants such as smoothed fictitious play \citep{ref:Fudenberg93} and the individual Q-learning \citep{ref:Leslie05} have been shown to converge equilibrium in the repeated play of two-agent zero-sum games (and beyond) if every agent follow the same dynamic with the same parameters \citep{ref:Robinson51,ref:Harris98,ref:Hofbauer05,ref:Leslie05}. 
Recently, certain extensions of these dynamics have been shown to converge equilibrium in two-agent zero-sum stochastic games, but again only if every agent follow the same dynamic with the same parameters \citep{ref:Leslie20,ref:Sayin20,ref:Sayin21}.

However, experimental studies on human-human interactions across the repeated play of games show considerable heterogeneity, e.g., in terms of how much they value past vs recent observations \citep{ref:Cheung97}. Hence, heterogeneous learning has received attention in the field of economics. For example, in \citep{ref:Giannitsarou03}, the authors examine the effects of heterogeneity, stemming from distinct priors, varying degrees of inertia in updates, and more, within the realm of macroeconomics, and reveals that homogeneous and heterogeneous learning dynamics yield different outcomes when analyzing local asymptotic stability. In \citep{ref:Michele07} and \citep{ref:Berardi12}, the authors investigate heterogeneity in learning processes by considering different information sets for forecasting, and examine the impact of heterogeneity on the speed of convergence of the learning process towards equilibrium. Our paper differs from these lines of works by providing a unifying framework to address heterogeneities for best-response-type dynamics. 

Heterogeneous learning has been studied extensively for cooperative multi-agent systems. Examples include distributed optimal coordination with heterogeneities induced by communication delays \citep{ref:Liu21}, distributed consensus optimization problems with heterogeneities induced by different computational capabilities \citep{ref:Wei22} and motion planning and robot swarms with heterogeneities induced by hardware, distinct perceptual and sensory capabilities, and the terrain of operation \citep{ref:Grabowski00, ref:Parker04, ref:Qu08, ref:Mathew15, ref:Spasojevic23}. However, due to their cooperative nature, the algorithms studied are not necessarily uncoupled and rational.

Heterogeneous learning in non-cooperative settings has recently received attention, but mainly for non-competitive settings, e.g., see \citep{ref:Wu23,ref:Yongacoglu23,ref:Wang23}. In \citep{ref:Wu23}, the authors addressed decentralized learning for Markov potential games through a two time-timescale learning algorithm where agents can use different learning rates. In \citep{ref:Yongacoglu23}, the authors addressed decentralized learning for stochastic games that induce weakly acyclic games under pure stationary strategy restriction through a two-phase algorithm where agents can use different lengths of exploration phases. In \citep{ref:Wang23}, the authors addressed learning in convex games where agents can have access to zeroth-order oracles or first-order gradient feedback. 

Heterogeneous learning in competitive settings has been studied mainly for different learning rates \citep{ref:Leslie03, ref:Chasnov20, ref:Daskalakis20}. In \citep{ref:Leslie03}, the authors addressed decentralized learning in two-agent zero-sum (and multi-agent identical-interest) normal-form games where agents learn at different timescales. In \citep{ref:Chasnov20}, the authors addressed gradient-based learning for continuous games where agents can use different learning rates. In \citep{ref:Daskalakis20}, the authors studied independent policy gradient algorithms for two-agent zero-sum stochastic games where agents learn at different timescales. The key difference is that in \citep{ref:Leslie03,ref:Daskalakis20}, the two-timescale learning plays an important role in proving convergence whereas here we show convergence even when agents use different step sizes. 

A preliminary version of the results of this paper appeared in \citep{ref:Sayin22d}. There we showed that the new variant of fictitious play for stochastic games, presented by \citep{ref:Sayin20}, converges to equilibrium in two-agent zero-sum stochastic games as long as the step sizes used by different agents are order-wise comparable and the discount factor is sufficiently small. Here, we analyze the convergence properties of any mixture of dynamics from an algorithm family addressing heterogeneities in terms of step sizes, exploration preferences, means in accessing to opponent actions and knowing the underlying model.

Notably, this paper is related to \citep{ref:Zhu10} and \citep{ref:Liu23}.
In \citep{ref:Zhu10}, the authors also studied heterogeneous learning in two-agent zero-sum SGs but mainly for different types of heterogeneities and only for the cases where agents have zero discount factor. Due to the zero discount factor, the agents only focus on immediate rewards, and therefore, stage games are stationary. In \citep{ref:Liu23}, the authors have recently addressed equilibrium seeking when agents' learning dynamics are constrained to local strategy set constraints and developed a distributed control algorithm guaranteeing convergence to generalized Nash equilibrium.

\textbf{Organization.} We provide preliminary information about SGs in Section \ref{sec:preliminary}. 
We introduce two families of best-response-type learning dynamics for repeated matrix games and SGs in Section \ref{sec:BR}. We characterize the convergence properties of any mixture of dynamics from these families for near zero-sum matrix games and SGs in Section \ref{sec:main}. Section \ref{sec:examples} includes numerical examples. We conclude the paper with some remarks in Section \ref{sec:conclusion}. Appendix includes preliminary information about stochastic approximation methods and the proof of one technical lemma.

\textbf{Notation.}
For any finite set $A$, let $|A|$ denote its number of elements. Let $\mathbb{I}_{\{P\}}\in \{0,1\}$ be the indicator function whether the proposition $P$ holds or not. For sets $A$ and $B$, let $B^A$ denote the space of measurable functions from $A$ to $B$.

\section{Game Formulation}\label{sec:preliminary}

Two-agent zero-sum SGs (ZSSGs) are generalizations of Markov Decision Processes (MDPs) to two-agent competitive environments. Let agents be indexed by $i=1,2$ and agents $i$ and $j$ denote, resp., the typical agent and the typical opponent. Then, we can characterize a ZSSG by a tuple $\mathcal{M}=\langle S,(A^i,r^i)_{i=1}^2,p,\gamma \rangle$. Here, $S$ denotes the finite set of states and $A^i$ denotes agent $i$'s finite set of actions at each state.\footnote{The generalization to state-variant action sets is rather straightforward.} 
The reward $r^i:S\times A^i\times A^j \rightarrow \mathbb{R}$ represents agent $i$'s stage payoff function. Due to its zero-sum nature, we have $r^1(s,a^1,a^2)+r^2(s,a^2,a^1)= 0$ for all $(s,a^i,a^j)$. Similarly, $p(s_+| s,a)$ for each $(s,a,s_+)$ denotes the transition probability from $s$ to $s_+$ when agents play the action profile $a=(a^i,a^j)$. Lastly, $\gamma \in [0,1)$ is the discount factor.

Agents take their actions to maximize the expected sum of discounted rewards they receive over infinite horizon. To this end, they can follow Markov stationary strategies that determine the probability of any action played depending only on the current state. For example, denote agent $i$'s Markov stationary strategy and the strategy space, resp., by $\pi^i:S\rightarrow \Delta^i$ and $\Pi^i$, where $\Delta^i$ is the probability simplex over $A^i$. Then, at every visit to $s$, agent $i$ plays action $a^i\sim\pi^i(s)$ independent of the past plays and the opponent play. 

If the opponent $j$ follows Markov stationary strategy $\pi^j$, then agent $i$ faces an MDP, for which there exists a Markov stationary solution. Without loss of generality, agent $i$'s goal is to find $\pi^i\in\Pi^i$ maximizing
\be\label{eq:value}
U^i(\pi^i,\pi^j) := \Exp\left[  \sum_{k=0}^{\infty} \gamma^k \; r^i(s_k,a_k) \right],
\ee
where $(s_k,a_k)$ is the pair of state and action profile at stage $k=0,1,\ldots$ and the expectation is taken with respect to the randomness on states visited and actions played when agents play according to the strategy profile $(\pi^i,\pi^j)$.
However, this objective is not a well-defined optimization problem due to its dependence on how the opponent plays. The following is a definition of Markov stationary equilibrium to predict the outcome of such non-cooperative interactions and Shapley proved its existence for ZSSGs in his inaugural paper \citep{ref:Shapley53}.

\begin{definition}{Markov Stationary Equilibrium}\label{def:NESG}
 A strategy profile $(\pi_*^1,\pi_*^2)\in \Pi^1\times \Pi^2$ is \textit{near Markov stationary equilibrium} of $\mathcal{M}$ with approximation error $\epsilon\geq 0$ provided that 
 $U^i(\pi^i_*,\pi_*^j) \geq U^i(\pi^i,\pi_*^j) - \epsilon$ for all $\pi^i\in \Pi^i$ and $i=1,2$. Call $(\pi_*^1,\pi_*^2)$ Markov stationary equilibrium if $\epsilon=0$.
\end{definition}

Note that for $|S|=1$ and $\gamma=0$, ZSSGs reduce to zero-sum matrix games (ZSGs) characterized by the tuple $\mathcal{G}=\langle A^i,R^i\rangle_{i=1}^2$, where $R^i\in \mathbb{R}^{|A^i|\times |A^j|}$ such that $(\pi^i)^TR^i\pi^j = \Exp_{(a^i,a^j)\sim(\pi^i,\pi^j)}[r^i(a^i,a^j)]$ for all $(\pi^i,\pi^j)$. View any action $a^i\in A^i$ as a pure strategy in which the associated action gets played with probability $1$, i.e., $A^i \subset \Delta^i$, and define $\BRe^i:\mathbb{R}^{|A^i|}\rightarrow 2^{\Delta^i}$ by 
\be\label{eq:BRe}
\BRe^i(q^i) := \{a^i\in A^i: (a^i)^Tq^i \geq (\tilde{a}^i)^Tq^i - \epsilon \;\forall \tilde{a}^i\in A^i\}.
\ee
Let $\BR^i(q^i):=\BRe^i(q^i)$ for $\epsilon=0$.
Then, we can define Nash equilibrium in $\mathcal{G}$ as follows.

\begin{definition}{Nash Equilibrium (NE)}\label{def:NE}
 A strategy profile $(\pi_*^1,\pi_*^2)\in \Delta^1\times \Delta^2$ is \textit{$\epsilon$-Nash equilibrium} of $\mathcal{G}$ provided that 
$\pi^i_*\in \BRe(R^i\pi^j_*)$ for $i=1,2$ and $j\neq i$. Call $(\pi_*^1,\pi_*^2)$ Nash equilibrium if $\epsilon = 0$. 
\end{definition}

Let $\sBR^i:\mathbb{R}^{|A^i|}\rightarrow\mathrm{int}(\Delta^i)$ denote the smoothed best response under entropy regularization, defined by\footnote{Given a set $A$, let $\mathrm{int}(A)$ denote its interior.}
\be\label{eq:sBR}
\sBR^i(q^i) := \argmax_{\mu^i\in\Delta^i}\big\{(\mu^i)^Tq^i + \tau^i H^i(\mu^i)\big\}
\ee
for all $q^i\in \mathbb{R}^{|A^i|}$, where $\tau^i>0$ is a temperature parameter determining the impact of the perturbation and the perturbation 
\be\label{eq:entropy}
H^i(\mu^i) := -\sum_{a^i} \mu^i(a^i)\log(\mu^i(a^i))\in [0,\log|A^i|].
\ee 
In the smoothed best response, any action gets played with some positive probability bounded from below by
\begin{align}\label{eq:lowerbound}
\sBR^i(q^i)(a^i) &=\frac{\exp(q^i(a^i)/\tau^i)}{\sum_{\tilde{a}^i\in A^i} \exp(q^i(\tilde{a}^i)/\tau^i)}\\
&\geq \exp(-2\|q^i\|_{\infty}/\tau^i)/|A^i| > 0\quad\forall a^i.
\end{align}

The following is a refinement of NE, also known as Nash distribution, to incorporate such bounded rationality \citep{ref:McKelvey95}.

\begin{definition}{Quantal Response Equilibrium (QRE)}\label{def:QRE}
 A strategy profile $(\pi_*^1,\pi_*^2)\in \Delta^1\times \Delta^2$ is \textit{quantal response equilibrium} of $\mathcal{G}$ provided that 
 $\pi^i_* = \sBR^i(R^i\pi^j_*)$ for $i=1,2$ and $j\neq i$. 
\end{definition}

Note that $\sBR^i(R^i\pi^j)$ has single value different from the set-valued function $\BRe^i(R^i\pi^j)$. Furthermore, we have $\sBR^i(R^i\pi^j)\in \BRe^i(R^i\pi^j)$ for $\epsilon \geq \tau^iH^i$. 

\section{A Family of Learning Dynamics}\label{sec:BR}

\begin{figure}[t!]
\centering
\includegraphics[width=.6\textwidth]{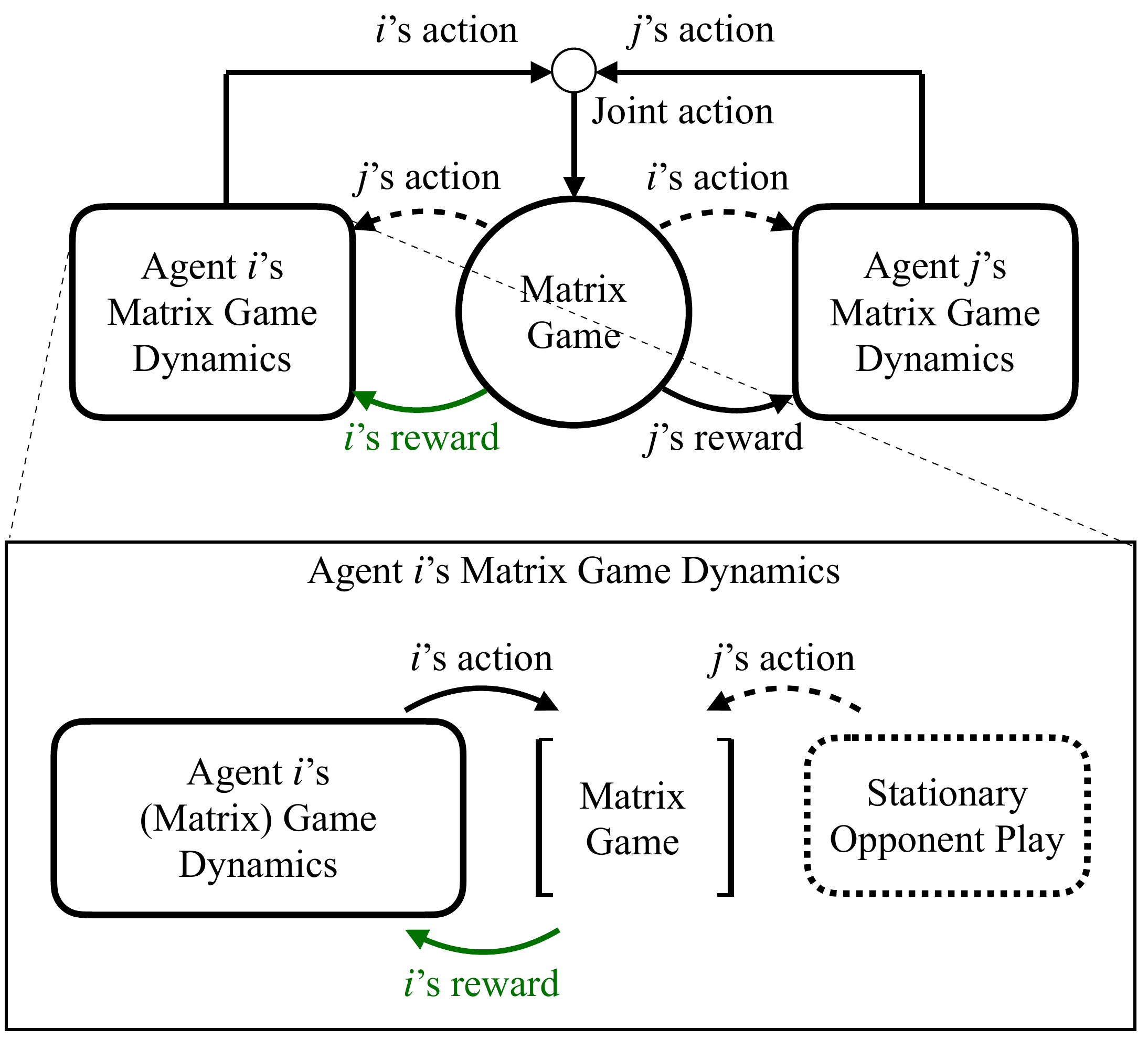}\caption{Agent $i$'s learning dynamics in the repeated play of a matrix game when the agent assumes that the opponent plays according to a stationary strategy. We use dashed arrows for opponent actions since agents may not have access to them.}\label{fig:repeated}
\end{figure}

This section describes two families of learning dynamics for the repeated play of matrix games and SGs.

\subsection{Algorithm Family \ref{tab:algoG} for Matrix Games}

Consider the repeated play of a matrix game $\mathcal{G}=\langle A^i,R^i\rangle_{i=1}^2$, as illustrated in Figure \ref{fig:repeated}. Each agent $i$ assumes that the opponent $j$ plays according to some underlying mixed strategy $\pi^j\in \Delta^j$. Then, the values of the local actions are given by $q^i=R^i\pi^j \in \mathbb{R}^{|A^i|}$, or equivalently $q^i(\cdot) = \Exp_{a^j\sim\pi^j}[r^i(\cdot,a^j)]\in \mathbb{R}^{A^i}$, with a slight abuse of notation since $A^i$ is a finite set. We call $q^i$ by \textit{local Q-function}. Agent $i$ can estimate $q^i$ based on the observations he/she makes throughout the repeated play of the underlying game. The observations can include the opponent's actions and the rewards received. Let $q_k^i$ denote the estimate at stage $k$. Then, agent $i$ responds greedily to the estimate $q_k^i$. 

As discussed above, the agents can have different means and interests, which can lead to heterogeneous learning dynamics. As a unifying framework, we present Algorithm Family \ref{tab:algoG}, in which algorithms are parameterized by the probability of access to opponent actions $\theta^i\in[0,1]$, the temperature parameter $\tau^i\geq 0$ and the step size $(\alpha^i_k)_{k\geq 0}$. The cases $\theta^i=1$ and $\theta^i=0$, resp., correspond to the dynamics where the agent has full access (i.e., belief-based) or no access (i.e., payoff-based) to the opponent's actions while the case $\theta^i\in (0,1)$ implies that the agent observes the opponent's action with probability $\theta^i$ at each stage independent of the history of the game. On the other hand, the case $\tau^i=0$ implies that the agent takes actions according to the best response $\BR^i(q_k^i)$, i.e., $a_k^i\in \BR^i(q_k^i)$, whereas the case $\tau^i > 0$ implies the agent responds according to the smoothed best response $\sBR^i(q_k^i)$, i.e., $a_k^i\sim\sBR^i(q_k^i)$. The latter can be preferred for exploration if $R^i$ is not known or for Hannan consistency, e.g., see \citep{ref:Fudenberg09}, though it can also lead to occasional mistakes.

\begin{algorithm}[tb]
\small
\caption{Heterogeneous Learning in ZSGs}
\begin{algorithmic}[1]\label{tab:algoG}
\REQUIRE $(\theta^i,\tau^i,(\alpha_k^i)_{k\geq 0})$
\STATE {\bfseries initialize:} $q_0^i$ arbitrarily
\FOR{each stage $k=0,1,\ldots$} 
\IF{$a_k^j$ can be observed and $R^i$ is known}
\STATE play $a_k^i\in\BR^i(q_k^i)$ or $a_k^i\sim\sBR^i(q_k^i)$ \hfill $\triangleright$ \textit{depending on $\tau^i$}
\STATE observe $a_k^j$ \hfill $\triangleright$ \textit{possible with probability $\theta^i$}
\STATE set $\hat{q}_k^i(\cdot) \equiv R^ia_k^j \in \mathbb{R}^{|A^i|}$
\STATE set $\bar{\alpha}_k^i(\tilde{a}^i) = \alpha_k^i$ for all $\tilde{a}^i\in A^i$
\ELSE
\STATE play $a_k^i\sim \sBR^i(q_k^i)$
\STATE observe $r_k^i = r^i(a_k^i,a_k^j)\in\mathbb{R}$
\STATE set $\hat{q}_k^i(\tilde{a}^i) = r_k^i$ for all $\tilde{a}^i\in A^i$
\STATE set $\bar{\alpha}_k^i(\tilde{a}^i) = \mathbb{I}_{\{\tilde{a}^i=a_k^i\}}\min\left\{1,\frac{\alpha_k^i}{\sBR^i(q_k^i)(\tilde{a}^i)}\right\}$\label{step:normalizationG}
\ENDIF
\STATE update $q_{k+1}^i(\cdot) = q_k^i (\cdot)+ \bar{\alpha}_k^i(\cdot)\big(\hat{q}_k^i(\cdot) - q_k^i(\cdot)\big)$\label{step:qupdate}
\ENDFOR
\end{algorithmic}
\end{algorithm}

Well-studied fictitious play \citep{ref:Brown51}, smooth fictitious play \citep{ref:Fudenberg93}, and individual Q-learning \citep{ref:Leslie05} are members of Algorithm Family \ref{tab:algoG}. For example, in fictitious play, agent $i$ forms a belief, say $\pi_k^j$, about the opponent's mixed strategy based on the opponent's actions observed. If the belief gets updated according to
\be\label{eq:belief}
\pi_{k+1}^j = \pi_k^j + \alpha_k^i(a_k^j - \pi_k^j)
\ee
with some step size $\alpha_k^i$, then $q_k^i = R^i\pi_k^j$ evolves as in step \ref{step:qupdate} in Algorithm Family \ref{tab:algoG}. On the other hand, in the individual Q-learning dynamics, agent $i$ normalizes the step size $\alpha_k^i$ with $\sBR^i(q_k^i)(a_k^i)$ to ensure that $q_k^i(a^i)$ for each $a^i$ gets updated at the same rate in the expectation and the threshold on the normalized step size ensures that the iterates remain bounded. However, the convergence guarantees for (smooth) fictitious play and individual Q-learning are for homogeneous cases where every agent follows the same dynamic with identical step sizes, e.g., see \citep{ref:Harris98,ref:Hofbauer02,ref:Leslie05}.

\subsection{Algorithm Family \ref{tab:algoSG} for Stochastic Games}

\begin{figure}[t!]
\centering
\includegraphics[width=.6\textwidth]{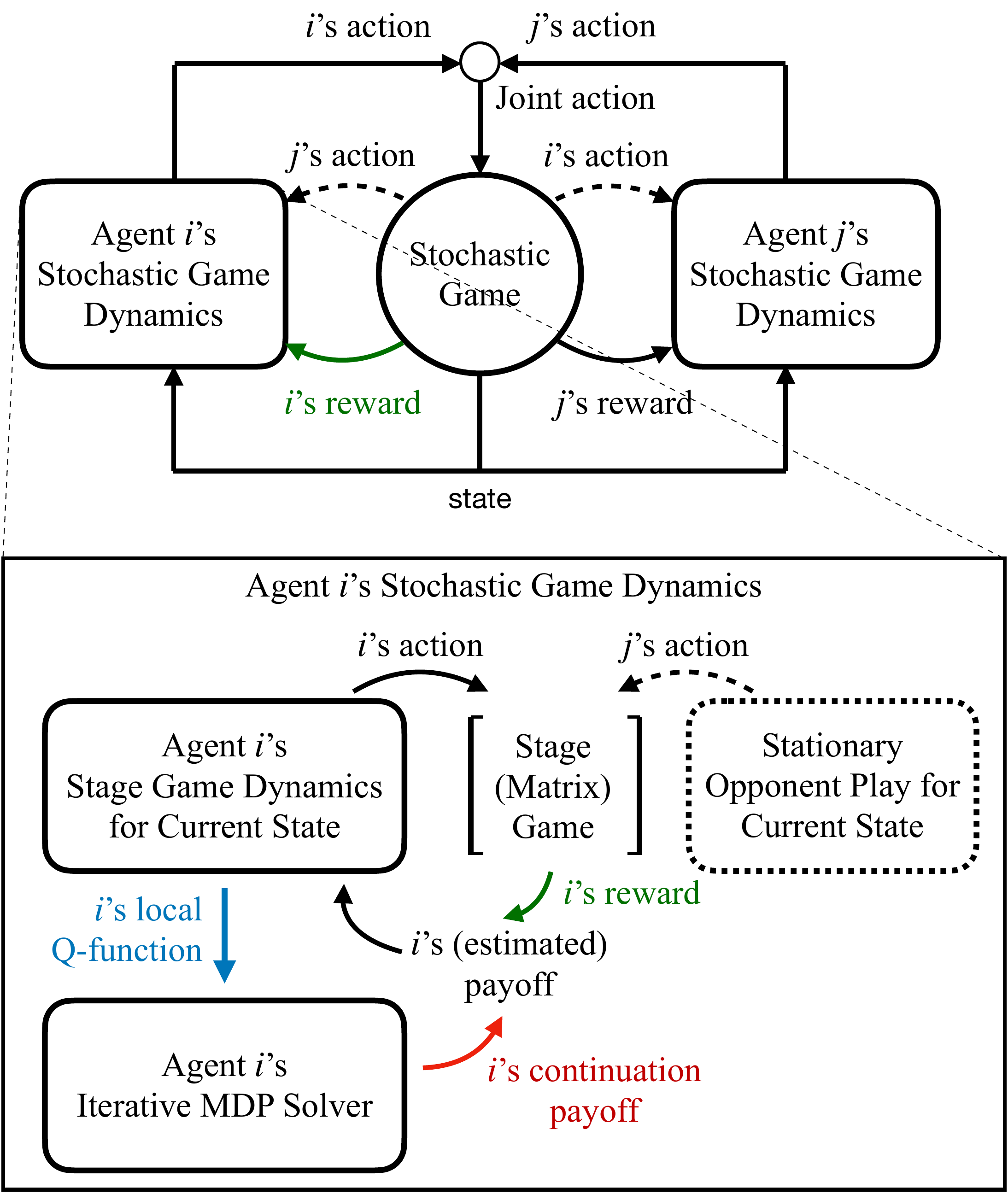}\caption{Agent $i$'s learning dynamics in SGs when the agent assumes that the opponent plays according to a Markov stationary strategy. We use dashed arrows for opponent actions since agents may not have access to them. The learning dynamics in SGs differ from the ones for repeated matrix games, illustrated in Figure \ref{fig:repeated} mainly due to the iterative MDP solver. Color coded arrows between stage game dynamics and the iterative MDP solver represent the coupling between them.}\label{fig:Markov}
\end{figure}

Consider learning in SGs. Following the trend in \citep{ref:Leslie20,ref:Sayin20,ref:Sayin21,ref:Wei21,ref:Baudin22,ref:Baudin22b}, we can extend Algorithm Family \ref{tab:algoG} to SG settings through the \textit{stage-game framework}. The stage-game framework was introduced by Shapley in \citep{ref:Shapley53}, and also used in Minimax-Q \citep{ref:Littman94} and Nash-Q \citep{ref:Hu03} algorithms. Particularly, given $\mathcal{M}=\langle S,(A^i,r^i)_{i=1}^2,p,\gamma\rangle$, we can view stage-wise interactions among agents as they are playing certain matrix games (called \textit{auxiliary stage games}) specific to each state whenever the associated state gets visited. Similar to the repeated play scheme, agents can follow an algorithm from Algorithm Family \ref{tab:algoG} for the current auxiliary stage game as if the opponent plays stationary strategy specific to the current state. In other words, agents can play the underlying auxiliary stage game as if the opponent plays according to Markov stationary strategy, as illustrated in Figure \ref{fig:Markov}. 
%

However, the payoffs of the auxiliary stage games depend both on the current reward received and the rewards to be received in the future stages, called \textit{continuation payoff}. 
Given that opponent $j$ follows Markov stationary strategy, say $\pi^j:S\rightarrow \Delta^j$, the principle of optimality yields that agent $i$'s payoffs in the auxiliary stage games, called \textit{global Q-function} and denoted by $Q^i:S\times A^i\times A^j\rightarrow\mathbb{R}$, are given by
\be\label{eq:Qfixedpoint}
Q^i(s,a) = r^i(s,a) + \gamma \sum_{s_+\in S} p(s_+\mid s,a) v^i(s_+)
\ee
for all $(s,a)$, and the value function $v^i:S\rightarrow\mathbb{R}$ is defined by
\be\label{eq:value}
v^i(s) = \max_{a^i\in A^i} \;\Exp_{a^j \sim \pi^j(s)} [Q^i(s,a^i,a^j)]
\ee
for all $s$. Under the stationary-opponent-play assumption, the agents can view that they play an underlying auxiliary stage game characterized by $\mathcal{G}(s) = \langle A^i,Q^i(s,\cdot)\rangle_{i=1}^2$ whenever $s$ gets visited. Correspondingly, their local actions in $\mathcal{G}(s)$ lead to the local Q-function 
$$q^i(s,\cdot) = \Exp_{a^j\sim\pi^j(s)}[Q^i(s,\cdot,a^j)]\in \mathbb{R}^{A^i}.$$ They can estimate their value functions and Q-functions, by solving \eqref{eq:Qfixedpoint} and \eqref{eq:value} iteratively, as illustrated in Figure \ref{fig:Markov}. 

\begin{algorithm}[t!]
\caption{Heterogeneous Learning for ZSSGs}
\small
\begin{algorithmic}[1]\label{tab:algoSG}
\REQUIRE $(\theta^i,\tau^i,(\alpha_t^i)_{t\geq 0},(\beta_t^i)_{t\geq 0})$
\STATE {\bfseries initialize:} $q_0^i$ and $v_0^i$ arbitrarily, and the counter $t^s_0 =0$, $\forall s$
\FOR{each stage $k=0,1,\ldots$}  
\STATE observe $s_k$

\IF{k=0} 
\STATE jump step \ref{step:play}
\ENDIF

\vspace{.1in}
\textit{Stage Game Dynamics for the Previous Stage Game:}
\STATE recall $s = s_{k-1}$ and $t = t_{k-1}^s$
\IF{$a_{t}^j(s)$ got observed and $(r^i,p(\cdot|\cdot))$ is known}
\STATE recall $a^j = a_t^j(s)$ and set 
$
\hat{q}^i_{t}(s,\cdot) \equiv r^i(s,\cdot,a^j) + \gamma \sum_{\tilde{s}\in S}p(\tilde{s}| s,\cdot,a^j) v_{t_{k-1}^{\tilde{s}}}^i(\tilde{s})
$
\STATE set $\bar{\alpha}_{t}^i(s,\tilde{a}^i) = \alpha_{t}^i$ for all $\tilde{a}^i\in A^i$
\ELSE
\STATE recall $r_{k-1}^i\in \mathbb{R}$ 
\STATE set $\hat{q}^i_{t}(s,\tilde{a}^i) = r_{k-1}^i + \gamma v_{t_{k-1}^{s_k}}^i(s_k)$ for all $\tilde{a}^i\in A^i$
\STATE recall $a^i =a_{t}^i(s)$ and set \label{step:normalization}
$
\bar{\alpha}_{t}^i(s,\tilde{a}^i)  = \mathbb{I}_{\{\tilde{a}^i=a^i\}}
\min\left\{1,\frac{\alpha_{t}^i}{\sBR^i(q_{t}^i(s,\cdot))(\tilde{a}^i)}\right\}
$
\ENDIF
\STATE update\label{step:qupdateSG} 
$
q_{t+1}^i(s,\cdot) \equiv q_{t}^i(s,\cdot)+\bar{\alpha}_{t}^i(s,\cdot)(\hat{q}^i_{t}(s,\cdot) - q_{t}^i(s,\cdot))
$

\vspace{.1in}
\textit{Iterative MDP Solver for the Previous Stage Game:}
\STATE set $\mu_t^i(s) = a^i_t(s)\in \BR^i\big(q_{t}^i(s,\cdot)\big)$ or $\mu_t^i(s) =\sBR^i\big(q_{t}^i(s,\cdot)\big)$
\STATE update \label{step:vupdateSG}
$
v_{t+1}^i(s) = v_{t}^i(s) + \beta_{t}^i\Big(\mu_t^i(s)^Tq_{t}^i(s,\cdot) - v_{t}^i(s)\Big)
$
\STATE increment the counter $t_{k}^{s'} = t_{k-1}^{s'}+\mathbb{I}_{\{s' = s\}}$ for all $s'\in S$

\vspace{.1in}
\textit{Play:}
\STATE recall $\tilde{s} = s_{k}$ and $\tilde{t} = t_{k}^{\tilde{s}}$
\IF{$a_k^j$ can be observed and $(r^i(\cdot),p(\cdot|\cdot))$ is known}\label{step:play}
\STATE play $a_{\tilde{t}}^i(\tilde{s})\in\BR^i\big(q_{\tilde{t}}^i(\tilde{s},\cdot)\big)$ or $a_{\tilde{t}}^i(\tilde{s})\sim\sBR^i\big(q_{\tilde{t}}^i(\tilde{s},\cdot)\big)$ \hfill $\triangleright$ \textit{depending on $\tau^i$}
\STATE observe $a_k^j$ \hfill $\triangleright$ \textit{possible with probability $\theta^i$}
\ELSE
\STATE play $a_{\tilde{t}}^i(\tilde{s})\sim\sBR^i\big(q_{\tilde{t}}^i(\tilde{s},\cdot)\big)$ 
\STATE observe $r_k^i = r^i(s_k^{},a_k^i,a_k^j)$
\ENDIF
\ENDFOR
\end{algorithmic}
\end{algorithm}

To address the heterogeneities that can arise due to the agents' different means and interests in a unifying framework, we present Algorithm Family \ref{tab:algoSG}, in which algorithms are parameterized by the probability of access to opponent actions $\theta^i\in [0,1]$, the temperature parameter $\tau^i\geq 0$, and the step size $(\alpha_t^i)_{t\geq 0}$ as in Algorithm Family \ref{tab:algoG}, and additionally the step size $(\beta_t^i)_{t\geq 0}$. The step size $\beta_t^i$ is used in the iterative MDP solver.

\begin{remark}
In Algorithm Family \ref{tab:algoG}, the updates of local $Q$ function estimates take place at the end. On the other hand, in Algorithm Family \ref{tab:algoSG}, the updates take place after the next state observed for a unified description addressing the one-stage look ahead for payoff-based learning. In other words, the update for the previous stage game takes place before the agents play in the current stage game.  
\end{remark}

\begin{remark}
The two-timescale model-based fictitious play dynamics presented by \citet{ref:Sayin20}, and the decentralized Q-learning dynamics presented by \citet{ref:Sayin21} can be viewed as members of Algorithm Family \ref{tab:algoSG}. However, the family does not include the two-timescale model-free fictitious play dynamics from \citep{ref:Sayin20}. The family instead treats model-free cases as if they are payoff-based. Agents by following this approach do not need to rely on the opponent to play every action with some positive probability to explore the underlying global Q-function properly. This can play an important role in exploration when the opponent prefers the best response rather than smoothed best response.
\end{remark}

\section{Convergence Results}\label{sec:main}

This section addresses the convergence properties of any mixture of learning dynamics from Algorithm Families \ref{tab:algoG} and \ref{tab:algoSG}, resp., for ZSGs and ZSSGs.

\subsection{Convergent heterogeneous Learning for ZSGs}

In Algorithm Family \ref{tab:algoG}, agents can use different step sizes satisfying the following standard assumption: 

\begin{assumption}\label{assume:step}
The local step sizes $(\alpha_k^i\in(0,1))_{k=0}^{\infty}$ decay monotonically $\alpha_k^i\rightarrow 0$ as $k\rightarrow \infty$ and $\sum_{k\geq 0} \alpha_k^i = \infty$ for each $i=1,2$. Furthermore, $\sum_{k\geq 0} (\alpha_k^i)^2 < \infty$ if agent $i$ plays according to the smoothed best response $\sBR^i(\cdot)$.
\end{assumption}

In the convergence analysis, we apply \textit{stochastic differential inclusion approximation} methods from \citep{ref:Benaim05} as the best response may not be unique. The preliminary information about these methods are provided in Appendix \ref{app:prelim}. For a unified representation, we let $\tau^i$ (or $\tau^j$) be zero if agent $i$ (or agent $j$) plays according to the best response, and define
\begin{subequations}\label{eq:val}
\begin{align}
&\maxmin^i(R^i):= \max_{\mu^i\in \Delta^i}\min_{\mu^j\in \Delta^j}\{(\mu^i)^TR^i\mu^j\}\\
&\val^i(R^i) := \max_{\tilde{\mu}^i\in\Delta^i}\min_{\tilde{\mu}^j\in \Delta^j} \Big\{(\tilde{\mu}^i)^TR^i\tilde{\mu}^j + \tau^i H^i(\tilde{\mu}^i) - \tau^j H^j(\tilde{\mu}^j)\Big\},
\end{align}
\end{subequations}
corresponding, resp., to the NE and QRE equilibrium values of the zero-sum game with payoff matrix $R^i$.\footnote{The uniqueness of the equilibrium value follows from the Minimax Theorem.} Even under perturbations, $\val^i(\cdot)$ has the contraction property with respect to the maximum norm, denoted by $\|\cdot\|_{\max}$, as shown in the following lemma, whose proof is moved to Appendix \ref{app:lemproofs}.

\begin{lemma}\label{lem:contraction}
For $\maxmin^i:\mathbb{R}^{|A^i|\times |A^j|}\rightarrow\mathbb{R}$ and $\val^i:\mathbb{R}^{|A^i|\times |A^j|}\rightarrow\mathbb{R}$, as described in \eqref{eq:val}, we have
\begin{align*}
&\underline{r} \leq |\maxmin^1(R^1) + \maxmin^2(R^2)| \leq \overline{r}\\
&\underline{r} \leq |\val^1(R^1) + \val^2(R^2)| \leq \overline{r},
\end{align*} 
where $\underline{r}$ and $\overline{r}\in\mathbb{R}$ are, resp., the minimum and maximum entries of the matrix $R^1+(R^2)^T$, and
$$
-\tau^j\log|A^j| \leq \val^i(R^i) - \maxmin^i(R^i)\leq \tau^i\log|A^i|
$$
for any $R^i\in \mathbb{R}^{|A^i|\times |A^j|}$. 
\end{lemma}


The following theorem characterizes the convergence properties of heterogeneous learning dynamics in near ZSGs.

\begin{theorem}\label{thm:G}
Consider the repeated play of $\mathcal{G}=\langle A^i,R^i\rangle_{i=1}^2$. Suppose that agents follow a learning dynamic from Algorithm Family \ref{tab:algoG} and the local step sizes satisfy Assumption \ref{assume:step}. Furthermore, the local step sizes are comparable such that $\lim_{k\rightarrow\infty}\alpha_k^i/\alpha_k^j\in(0,\infty)$ and, without loss of generality, we assume that $\alpha_k^1/\alpha_k^2 \rightarrow d \in (0,1]$ as $k\rightarrow\infty$.
Let $\pi_k^i\in\Delta^i$ denote a \textit{weighted empirical average} of the agent $i$'s past actions until stage $k$. Let $a^i_k\in A^i$ denote the action played by agent $i$ at stage $k$. Then, $\pi_k^i$ evolves according to 
\be\label{eq:belief}
\pi_{k+1}^i = \pi_k^i + \alpha_{k}^j(a^i_k - \pi_k^i)
\ee
with arbitrary $\pi_0^i$. We have $\|q_k^i-R^i\pi_k^j\|_2\rightarrow 0$ as $k\rightarrow \infty$ and
\be\label{eq:limsup}
\limsup_{k\rightarrow\infty} (\delta_k^1 + d\cdot\delta_k^2) \leq c
\ee
almost surely, where
\begin{align}
&\delta^i_k := (\mu_k^i)^Tq_k^i + \tau^i H^i(\mu_k^i) - \tau^j H^j(\pi_k^j) - \val^i(R^i),\label{eq:delta}\\
&c := \lambda \|R^1+(R^2)^T\|_{\max} - (\val^1(R^1)+\val^2(R^2))\label{eq:c}
\end{align}
for some $\lambda > 1$, and $\mu_k^j = a_k^j\in \BR^j(q_k^j)$ (or $\mu_k^j = \sBR^j(q_k^j)$) if agent $j$ plays according to the best response (or smoothed best response). We also have $\delta^i_k\geq\|q_k^i-R^i\pi_k^j\|_2$.
\end{theorem}

A sketch of the proof is as follows: We first approximate the discrete-time update of local Q-function estimates and empirical averages of actions via continuous-time differential equations (or inclusions) such that the limit set of the former is included in the compact, connected, internally chain transitive set of the latter. Then, we formulate a unifying Lyapunov function for the latter to characterize the compact, connected, internally chain transitive set, and therefore, the limit set of the former.

\begin{myproof}
%
The local Q-function estimates remain bounded in Algorithm Family \ref{tab:algoG} since $\bar{\alpha}_k^i(a^i)\in[0,1]$ for all $a^i\in A^i$. For example, $\limsup_{k\rightarrow\infty} \|q_k^i\|_{\infty} \leq \|R^i\|_{\max}$. By \eqref{eq:lowerbound}, the boundedness of the estimates and decaying step size imply that the normalization at step \ref{step:normalizationG} becomes redundant eventually. Hence, for any $\epsilon>0$, there exists a stage $K_{\epsilon}$ such that $\bar{\alpha}_k^i(a^i) = \mathbb{I}_{\{a^i=a_k^i\}} \alpha_k^i/\sBR^i(q_k^i)(a^i)$ and $q_k^i \in \Xi^i$, where $\Xi^i$ is a compact subset of $\mathbb{R}^{|A^i|}$, for all $k\geq K_{\epsilon}$. The evolution of the local Q-function estimates $q_k^i$ and the empirical averages of actions $\pi_k^j$ can be written as 
\begin{align}\label{eq:DiffInc}
\begin{bmatrix}
q_{k+1}^i \\ \pi_{k+1}^j 
\end{bmatrix} = 
\begin{bmatrix}
q_{k}^i \\ \pi_{k}^j 
\end{bmatrix} +  \alpha_k^i \left(\begin{bmatrix}
R^i\mu_k^j \\ \mu_k^j 
\end{bmatrix} + \begin{bmatrix}\uw_{k}^i \\ \ow_{k}^j  \end{bmatrix} -\begin{bmatrix}
q_{k}^i \\ \pi_{k}^j 
\end{bmatrix}\right)
\end{align}
for all $k\geq K_{\epsilon}$. In the update of $\pi^j_k$, the noise term $\ow_k^j = a_k^j - \mu_k^j$. In the update of $q_k^i$, the noise term $\uw_k^i = R^ia_k^j - R^i\mu_k^j$ if agent $i$ observes $a_k^j$ and knows $R^i$. If agent $i$ does not observe $a_k^j$ or know $R^i$, then we have
\begin{align}\label{eq:uw}
\uw_k^i(a^i) = &\;\frac{\mathbb{I}_{\{a^i = a_k^i\}}}{\mu_k^i(a^i)}(r^i(a^i,a_k^j) - q_k^i(a^i))- \big(\Exp_{a^{j}\sim\mu_k^j}[r^i(a^i,a^j)]- q_k^i(a^i)\big). 
\end{align}

Let $x_k = [q_k^i;\pi_k^j]_{i=1}^2$. If agent $j$ plays according to the best response, let $F^{ij}(x_k)\subset \mathbb{R}^{|A^i|+|A^j|}$ be a set-valued function defined by\footnote{Let $\mathrm{Conv}(X)$ denote the convex hull of the set $X$.}
\be\label{eq:Fijbest}
F^{ij}(x_k) = \left\{\begin{bmatrix}
R^i\mu^j \\ \mu^j 
\end{bmatrix} -\begin{bmatrix}
q_{k}^i \\ \pi_{k}^j 
\end{bmatrix}: \mu^j \in \mathrm{Conv}(\BR^j(q_k^j))\right\}.
\ee
If agent $j$ plays according to the smoothed best response, let $F^{ij}(x_k) \in\mathbb{R}^{|A^i|+|A^j|}$ be a single-valued function defined by
\be\label{eq:Fijsmooth}
F^{ij}(x_k) = \begin{bmatrix}
R^i\sBR^j(q_k^j) \\ \sBR^j(q_k^j) 
\end{bmatrix} -\begin{bmatrix}
q_{k}^i \\ \pi_{k}^j 
\end{bmatrix}.
\ee
Note that the noise-free innovation term in \eqref{eq:DiffInc} satisfies $\begin{bmatrix}
R^i\mu_k^j \\ \mu_k^j 
\end{bmatrix} -\begin{bmatrix}
q_{k}^i \\ \pi_{k}^j 
\end{bmatrix} \in F^{ij}(x_k)$. Therefore,  the evolution of the local Q-function estimates $q_k^i$ and empirical averages of actions $\pi_k^j$ satisfies
\begin{align}
\begin{bmatrix}
q_{k+1}^i \\ \pi_{k+1}^j 
\end{bmatrix} - 
\begin{bmatrix}
q_{k}^i \\ \pi_{k}^j 
\end{bmatrix} - \alpha_k^i\begin{bmatrix}\uw_{k}^i \\ \ow_{k}^j  \end{bmatrix} 
\in \alpha_k^i \cdot F^{ij}(x_k).
\end{align}
 
We can show that $x_k\in X = \prod_{i=1}^2 \Xi^i \times \Delta^j \subset \mathbb{R}^{2m}$, where $m=|A^1|+|A^2|$, satisfies
\be\label{eq:x}
x_{k+1} - x_k - \alpha_k (w_k+e_k) \in \alpha_k \cdot F(x_k),
\ee
for all $k\geq K_{\epsilon}$, where the step size $\alpha_k := \alpha_k^1$, the noise term $w_{k} = \Big[\uw_{k}^1;\ow_{k}^2;\frac{\alpha_k^2}{\alpha_k^1}\uw_k^2;\frac{\alpha_k^2}{\alpha_k^1}\ow_k^1\Big]$, the set-valued map $F(x_k) \subset \mathbb{R}^{2m}$ is given by
\begin{align}\label{eq:F}
F(x_k) = \bigg\{\begin{bmatrix}f^{12} \\ \frac{1}{d} \cdot f^{21} \end{bmatrix}: &\;f^{ij}\in F^{ij}(x_k) \mbox{ for } i =1,2 \bigg\},
\end{align}
and the error term is defined by
\be
e_k = \begin{bmatrix}\mathbf{0} \\ \left(\frac{\alpha_k^2}{\alpha_k^1} - \frac{1}{d}\right)\left(\begin{bmatrix}
R^2\mu_k^1 \\ \mu_k^1 
\end{bmatrix} -\begin{bmatrix}
q_{k}^2 \\ \pi_{k}^1 
\end{bmatrix} \right) \end{bmatrix},
\ee
which is asymptotically negligible as $\alpha_k^2/\alpha_k^1 \rightarrow 1/d \in [1,\infty)$.

For the cases where agent $j$ plays according to the best response, the set-valued function $F^{ij}(\cdot)$, as described in \eqref{eq:Fijbest}, is a Marchaud map (described in Appendix \ref{app:prelimContinuous}) because the best response function $\BR^j(\cdot)$ is a Marchaud map, e.g., see \citep[Section 2.2]{ref:Benaim05}. For the cases where agent $j$ plays according to the smoothed best response, the single-valued function $F^{ij}(\cdot)$, as described in \eqref{eq:Fijsmooth}, is Lipschitz continuous since the smoothed best response function $\sBR^j(\cdot)$ is  $1/\tau$-Lipschitz continuous. Since $F^{ij}(\cdot)$ is a Marchaud map in each case, by \eqref{eq:F},  we can show that $F(\cdot)$ is a Marchaud map.

Let $(\Omega,\mathcal{F},\mathbb{P})$ be the underlying probability space and 
$\{\mathcal{F}_k = \sigma(q_l^i,\pi_l^i \mbox{ for } i=1,2, \mbox{ and } l\leq k)\}_{k\geq 0}$ be a filtration of $\mathcal{F}$. Note that $\uw_{k-1}^i$ and $\ow_{k-1}^i$ for each $i=1,2$, and therefore, $w_{k-1}$ are $\mathcal{F}_k$-measurable for each $k>0$. Furthermore, we have $\Exp[\ow_k^i | \mathcal{F}_k] = \mathbf{0}$ and $\Exp[\|\ow_k^i\|^2|\mathcal{F}_k]$ is bounded from above by some finite number uniformly for all $k\geq K_{\epsilon}$ since 
\begin{itemize}
\item $\ow_k^i = \mathbf{0}$ if agent $i$ plays according to the best response,
\item $\ow_k^i = a_k^i - \sBR^i(q_k^i)$, where $a_k^i\sim \sBR^i(q_k^i)$ with finite support, if agent $i$ plays according to the smoothed best response.
\end{itemize} 
On the other hand, if agent $i$ does not know $R^i$, then $\uw_k^i$ is as described in \eqref{eq:uw}, and therefore, we can show that $\Exp[\uw_k^i | \mathcal{F}_k] = \mathbf{0}$ and $\Exp[\|\uw_k^i\|^2|\mathcal{F}_k]$ is bounded from above by some finite number uniformly for all $k\geq K_{\epsilon}$. If agent $i$ knows $R^i$, then $\uw_k^i = R^ia_k^j - R^i \mu_k^j$ with probability $\theta^i$ and $\uw_k^i$ is as described in \eqref{eq:uw} with probability $(1-\theta^i)$ independent of $\mathcal{F}_k$. Therefore, we also have $\Exp[\uw_k^i | \mathcal{F}_k] = \mathbf{0}$ and $\Exp[\|\uw_k^i\|^2|\mathcal{F}_k]$ is bounded from above by some finite number uniformly for all $k\geq K_{\epsilon}$.

Correspondingly, $\Exp[w_k^i | \mathcal{F}_k] = \mathbf{0}$ and $\Exp[\|w_k^i\|^2|\mathcal{F}_k] < W$ for some finite $W$ for all $k\geq K_{\epsilon}$ since $w_k$ is a linear function of $\uw_k^i$'s and $\ow_k^i$'s while the deterministic $\alpha_k^2/\alpha_k^1$ is uniformly bounded away from zero for all $k\geq K_{\epsilon}$. 
Furthermore, Assumption \ref{assume:step} holds for the step size $\alpha_k$. Therefore, the iterate $x_k$ converges to a compact connected internally chain transitive set of the differential inclusion $\dot{x} \in F(x)$. 

Since $F(\cdot)$ is a Marchaud map, there always exists a solution to $\dot{x} \in F(x)$ with initial point $x_o \in X$ that is an absolutely continuous mapping $x:\mathbb{R}\rightarrow X$ such that $x(0)= x_o$, and $\frac{dx(t)}{dt} \in F(x(t))$ for almost every $t\in \mathbb{R}$ \citep{ref:Benaim05}. Define $q^i:\mathbb{R}\rightarrow \Xi^i$ and $\pi^j:\mathbb{R}\rightarrow \Delta^j$ such that $x(t) = [q^i(t);\pi^j(t)]_{i=1}^2$. Then, for almost every $t\in \mathbb{R}$, $q^i(t)$ and $\pi^j(t)$ satisfy
\begin{subequations}\label{eq:DIG}
\begin{align}
&\frac{dq^i(t)}{dt} = d^i(R^i \mu^j(t) - q^i(t))\\
&\frac{d\pi^j(t)}{dt} = d^i(\mu^j(t) - \pi^j(t))
\end{align}
\end{subequations}
for each $i=1,2$ and $j\neq i$, where $\mu^j(t) \in \BR^j(q^j(t))$ (and $\mu^j(t) = \sBR^j(q^j(t)))$ if agent $j$ plays according to the best response (and the smoothed best response), and $d^1 = 1$ and $d^2 = 1/d \in [1,\infty)$.

Next, we define the continuous function $V:X \rightarrow [0,\infty)$ by\footnote{Let $(x)_+ := (x + |x|)/2$.}
\begin{align}
V(q^1,\pi^2,q^2,\pi^1) = &\;\left(L^1(q^1,\pi^2) + d \cdot L^2(q^2,\pi^1) - c\right)_++ \sum_{i=1}^2 \|q^i - R^i\pi^j\|_2,\label{eq:V}
\end{align}
where $c$ is as described in \eqref{eq:c} and $L^i:\Xi^i\times \Delta^j\rightarrow \mathbb{R}$ is defined by
\begin{align}\label{eq:L}
L^i(q^i,\pi^j) := &\; \max_{\tilde{\mu}^i\in \Delta^i} \left\{(\tilde{\mu}^i)^Tq^i + \tau^i H^i(\tilde{\mu}^i) - \tau^j H^j(\pi^j)\right\}+ \|q^i - R^i\pi^j\|_2 - \val^i(R^i).
\end{align}
Lemma \ref{lem:contraction} yields that $c \geq (\lambda -1)\|R^1+(R^2)^T\|_{\max}\geq 0$ and $c=0$ if and only if $\|R^1+(R^2)^T\|_{\max}=0$ as $\lambda > 1$.

Let $\bar{\mu}^i \in \argmax_{\tilde{\mu}^i}\left\{(\tilde{\mu}^i)^TR^i\pi^j + \tau^i H^i(\tilde{\mu}^i)\right\}$.\footnote{We have set inclusion rather than equality to address the case $\tau^i=0$.} Then, we have
\begin{align}
L^i(q^i,\pi^j) \geq &\;(\bar{\mu}^i)^Tq^i + \tau^iH^i(\bar{\mu}^i) - \tau^j H^j(\pi^j) + \|q^i-R^i\pi^j\| - \val^i(R^i)\nn\\
=&\; \underbrace{(\bar{\mu}^i)^TR^i\pi^j + \tau^iH^i(\bar{\mu}^i) - \tau^j H^j(\pi^j) - \val^i(R^i)}_{\geq 0}+\underbrace{(\bar{\mu}^i)^T(q^i - R^i\pi^j) + \|q^i-R^i\pi^j\|_2}_{\geq 0}
\end{align}
where the first term on the right-hand side is non-negative by the definitions of $\val^i(R^i)$ and $\bar{\mu}^i$, and the second term is non-negative as $\bar{\mu}^i\in\Delta^i$ is a probability distribution and $\|x\|_2\geq \|x\|_{\infty}$ for any vector $x$. Hence, $L^i(q^i,\pi^j)\geq 0$ is a \textit{non-negative} function.

\begin{remark}
If the underlying game is zero-sum, then we have $c = 0$ and $(L^1+d L^2)_+ = L^1+dL^2$. Furthermore, if we also had $q^i = R^i\pi^j$ and $d=1$, then the function \eqref{eq:V} reduces to 
\be
\sum_{i=1}^2\max_{\tilde{\mu}^i\in \Delta^i} \left\{(\tilde{\mu}^i)^TR^i\pi^j + \tau^i H^i(\tilde{\mu}^i) - \tau^j H^j(\pi^j)\right\},
\ee
as formulated in \citep{ref:Harris98} (and \citep{ref:Hofbauer05}) if both agents play according to the best response with $\tau^1=\tau^2 = 0$ (and the smoothed best response with $\tau^1,\tau^2>0$).

The candidate function \eqref{eq:V} also resembles the Lyapunov functions formulated in \citep{ref:Sayin21} and \citep{ref:Sayin22d}. The function differs from the former by addressing asymmetric step sizes and convergence to quantal response equilibrium when agents play according to the smoothed best response. The function differs from the latter by addressing the cases with smoothed best response and payoff-based learning where $q^i$ may not be equal to $R^i\pi^j$. In that respect, we can view \eqref{eq:V} as a unifying Lyapunov function addressing all these cases.
\end{remark}

In the following, we show the validity of \eqref{eq:V} as a Lyapunov function for \eqref{eq:DIG}. Given a solution $x(t) = [q^i(t);\pi^j(t)]_{i=1}^2$ to \eqref{eq:DIG}, the time derivatives of $\|q^i-R^i\pi^j\|$ and $L^i(q^i(t),\pi^j(t))$ are, resp. given by
\begin{subequations}
\begin{align}
&\frac{d\|q^i-R^i\pi^j\|}{dt} = \frac{(q^i-R^i\pi^j)^T(\dot{q}^i-R^i\dot{\pi}^j)}{\|q^i-R^i\pi^j\|}\\
&\frac{dL^i}{dt} = (\mu^i)^T\dot{q}^i - \tau^j \nabla H^j(\pi^j)\dot{\pi}^j + \frac{d\|q^i-R^i\pi^j\|}{dt}
\end{align}
\end{subequations}
almost everywhere, based on \citep{ref:Harris98} (and \citep{ref:Hofbauer05}) if agent $i$ plays according to the best response (and the smoothed best response). After some algebra, we obtain $\frac{d}{dt}\|q^i-R^i\pi^j\| = -d^i\|q^i-R^i\pi^j\|$ and 
\begin{align}
\frac{dL^i}{dt} \leq &\;d^i\left[(\mu^i)^TR^i\mu^j -\val^i + \tau^iH^i(\mu^i)-\tau^jH^j(\mu^j)\right]-d^i\cdot L^i(q^i,\pi^j),
\end{align}
where the inequality follows from $H^j(\mu^j)-H^j(\pi^j) - \nabla H^j(\pi^j)(\mu^j-\pi^j) \leq 0$ due to the concavity of entropy function. Recall that $d^1=1$ and $d^2 = 1/d$. Therefore, we have
\begin{align}
\frac{d}{dt}&(L^1+d\cdot L^2) \leq (\mu^1)^T(R^1+(R^2)^T)\mu^2 - (\val^1(R^1)+\val^2(R^2)) - L^1 - d\cdot L^2 +\underbrace{(d-1)L_2}_{\leq 0},\nn
\end{align}
where the last term on the right-hand side is non-positive since $d\in(0,1]$ and $L^2(\cdot)$ is a non-negative function. By \eqref{eq:c}, we obtain
\be
\frac{d}{dt}(L^1+d\cdot L^2) \leq -(L_1+d\cdot L_2 - c),
\ee
with strict inequality in non-zero-sum games, i.e., when $\|R^1+(R^2)^T\|_{\max}\neq 0$. Hence, the set $\Lambda = \{(q^1,\pi^2,q^2,\pi^1)\in X: L_1(q^1,\pi^2)+dL_2(q^2,\pi^1) \leq  c \mbox{ and } q^i=R^i\pi^j,\mbox{ for }i=1,2\}$ is positively invariant for almost every solution of \eqref{eq:DIG} and $V(\cdot)$ is a Lyapunov function for $\Lambda$. Note that $V(\Lambda) = 0$ is a singleton. Therefore, we have
\be
\lim_{k\rightarrow \infty} V(q_k^1,\pi_k^2,q_k^2,\pi_k^1) = 0
\ee
almost surely. Then, the definition of $V(\cdot)$ implies that there exists an asymptotically negligible sequence $\zeta_k\rightarrow 0$ such that
\be\label{eq:zeta}
L^1(q_k^1,\pi_k^2) + d \cdot L^2(q^2_k,\pi_k^1) \leq c + \zeta_k\quad\forall k,
\ee
and $\|q_k^i - R^i\pi_k^j\|_2\rightarrow 0$ almost surely for each $i=1,2$. By \eqref{eq:L} and the non-negativity of $L^i(\cdot)$, we have
\begin{align}\nn
\delta_k^1 + d\cdot\delta_k^2\leq c + \zeta_k -\|q^1_k-R^1\pi_k^2\|_2 - d\cdot \|q^2_k - R^2\pi_k^1\|_2,
\end{align}
and $\delta_k^i \geq \|q^i_k-R^i\pi_k^j\|_2$, where $\delta_k^i$ is as described in \eqref{eq:delta}. The proof is completed as $\|q_k^i - R^i\pi_k^j\|_2\rightarrow 0$.
\end{myproof}

The following corollary to Theorem \ref{thm:G} characterizes the near equilibrium convergence of the dynamics in near ZSGs. To this end, we define
\be
u^i(\pi^i,\pi^j) := (\pi^i)^TR^i\pi^j + \tau^iH^i(\pi^i) - \tau^j H^j(\pi^j)
\ee
for all $(\pi^i,\pi^j)\in \Delta^i\times \Delta^j$ and $i=1,2$.

\begin{corollary}\label{cor:G}
Consider the repeated play of $\mathcal{G}=\langle A^i,R^i\rangle_{i=1}^2$. Suppose that agents follow a learning dynamic from Algorithm Family \ref{tab:algoG} and the local step sizes satisfy Assumption \ref{assume:step}. Furthermore, the local step sizes are comparable such that $\lim_{k\rightarrow\infty}\alpha_k^i/\alpha_k^j\in(0,\infty)$ and, without loss of generality, we assume that $\alpha_k^1/\alpha_k^2 \rightarrow d \in (0,1]$ as $k\rightarrow\infty$.
Let $\pi_k^i\in\Delta^i$ denote a \textit{weighted empirical average} of the agent $i$'s past actions, evolving according to \eqref{eq:belief} for all $k\geq 0$ with arbitrary $\pi_0^i$. Then, we have
\begin{align*}
&\limsup_{k\rightarrow \infty}\left(u^i(\pi^i_k,\pi^j_k) - \val^i(R^i)\right) \leq \frac{2}{d^i}\|\bar{R}\|_{\max}\\
&\liminf_{k\rightarrow \infty}\left(u^i(\pi^i_k,\pi^j_k) - \val^i(R^i)\right) \geq -2\left(1+\frac{1}{d^j}\right)\|\bar{R}\|_{\max}
\end{align*}
almost surely for each $i=1,2$ and $j\neq i$, where $\bar{R} := R^1+(R^2)^T$ and $(d^1,d^2)=(1,d)$ . Hence, $(\pi_k^1,\pi_k^2)$ converges almost surely to NE or QRE (depending on, resp., whether $\tau^i = 0$ or $\tau^i> 0$) in ZSGs, i.e., when $\|\bar{R}\|_{\max}=0$, for any $d\in(0,1]$. 
\end{corollary}

\begin{myproof}
Since $\delta_k^i\geq \|q_k^i-R^i\pi_k^j\|_2$ and $\|q_k^i-R^i\pi_k^j\|_2\rightarrow 0$, Theorem \ref{thm:G} yields that
\be\label{eq:uval}
u^i(\pi_k^i,\pi_k^j) - \val^i(R^i) \leq \frac{c}{d^i} + \zeta_k
\ee
for all $k\geq 0$ and $i=1,2$, for some $\zeta_k\rightarrow0$ as $k\rightarrow \infty$ almost surely. We have the upper bound as Lemma \ref{lem:contraction} yields that $c \leq (\lambda + 1)\|\bar{R}\|_{\max}$.

To compute a lower bound, we add and subtract $(\pi_k^i)^T(R^j)^T\pi_k^j$ to the left-hand side of \eqref{eq:uval}. Then, we can write the inequality as
\begin{align}
-(u^j(\pi_k^j,\pi_k^i) - \val^j&(R^j)) \leq \frac{c}{d^i} +\zeta_k - (\pi_k^i)^T\bar{R}\pi_k^j + \sum_{t=1}^2 \val^t(R^t).
\end{align}
Note that $(\pi_k^i)^T\bar{R}\pi_k^j\geq - \|\bar{R}\|_{\max}$ as $\pi_k^i$'s are probability distributions. Then, Lemma \ref{lem:contraction} yields that
\be
(u^j(\pi_k^j,\pi_k^i) - \val^j(R^j)) \geq -\frac{c}{d^i} - 2\|\bar{R}\|_{\max}-\zeta_k.
\ee
By changing the agent index from $j$ to $i$, we obtain the lower bound.
\end{myproof}

The following corollary to Theorem \ref{thm:G} shows the rationality of the dynamics from Algorithm Family \ref{tab:algoG}.

\begin{corollary}\label{cor:rationalG}
Consider the repeated play of $\mathcal{G}=\langle A^i,R^i\rangle_{i=1}^2$. Suppose that agent $i$ follows a learning dynamic from Algorithm Family \ref{tab:algoG} while the opponent $j$ plays according to mixed strategy $\pi^j\in \Delta^j$. The local step size $(\alpha_k^i)_{k\geq 0}$ satisfy Assumption \ref{assume:step}. Let $\pi_k^i\in\Delta^i$ denote a \textit{weighted empirical average} of the agent $i$'s past actions, evolving according to \eqref{eq:belief}  with arbitrary initialization. Then, we have
$u^i(\pi^i_k,\pi^j) \rightarrow \max_{\pi^i\in\Delta^i}\{u^i(\pi^i,\pi^j)\}$
almost surely as $k\rightarrow\infty$. 
\end{corollary}

\begin{myproof}
We can consider that agent $i$ is playing a ZSG against an opponent (other than agent $j$) following the same dynamic with a singleton action space and view agent $j$ as the Nature. Then, the proof follows from Corollary \ref{cor:G}.
\end{myproof}

\subsection{Convergent heterogeneous Learning for ZSSGs}

Algorithm Family \ref{tab:algoSG} involves two local step sizes $\alpha_k^i$ and $\beta_k^i$. The following is the counterpart of Assumption \ref{assume:step} for $\beta_k^i$.

\begin{assumption}\label{assume:stepbeta}
The local step sizes $(\beta_k^i\in(0,1))_{k=0}^{\infty}$ decay monotonically $\beta_k^i\rightarrow 0$ as $k\rightarrow \infty$ and $\sum_{k\geq 0} \beta_k^i = \infty$ for each $i=1,2$.
\end{assumption}

Furthermore, we assume that $\alpha_k^i$ and $\beta_k^i$ decay in different time scales at certain rates as in \citep[Assumption 1-ii]{ref:Sayin21}. 

\begin{assumption}\label{assume:stepbetacompare}
  Given any $M \in (0, 1)$, there exists a non-decreasing polynomial function $C(\cdot)$ (which may depend on M) such that  we have 
  \begin{equation}
\max \left\{ \ell \in \mathbb{Z}_+ | \ell \leq c \textrm{ and } \beta_\ell/\alpha_c > \lambda \right\} \leq Mc
\end{equation} 
for all $c \geq C(\lambda^{-1})$ for any $\lambda \in (0, 1)$ provided that $\left\{ \ell \in \mathbb{Z}_+ | \ell \leq c \textrm{ and } \beta_\ell/\alpha_c > \lambda \right\} \neq \emptyset$.
  \end{assumption}

In SGs, agents play stage games associated with the state visited. Whether each state gets visited infinitely often plays an important role in ensuring that agents play stage games associated with each state sufficiently many times to reach equilibrium. To this end, consider the directed graph $G$ whose vertices correspond to the states of the underlying SG and there is an edge from $s$ to $s'$ provided that 
\begin{itemize}
\item \textit{Case $i)$ - Both agent play according to the best response:} $p(s'\mid s,a) > 0 $ for all $a\in A$,
\item \textit{Case $ii)$ - Only agent $i$ plays according to the best response:} $p(s'\mid s, a^i,a^j) > 0$ for all $a^i\in A^i$ and at least for one $a^j\in A^j$,
\item \textit{Case $iii)$ - Both agent play according to the smoothed best response:} $p(s'\mid s,a) > 0$ at least for one action profile $a\in A$.
\end{itemize}
We make the following assumption for the graph $G$.

\begin{assumption}\label{assume:state}
The graph $G$ is strongly connected (each vertex is reachable from every other vertex).
\end{assumption}

Let $Q_*^i(s,a)$ and $v_*^i(s)$ denote the unique Q-function and value function associated with Markov stationary equilibrium of the underlying ZSSG. Then, they satisfy
\be\label{eq:Qstar}
Q_*^i(s,a) = r^i(s,a) + \gamma\sum_{s_+\in S}p(s_+\mid s,a)v_*^i(s_+)
\ee 
and $v_*^i(s) = \maxmin^i(Q_*^i(s,\cdot))$. We also have $v_*^1(s)+v_*^2(s) = 0$ for all $s$ in ZSSGs.

The following theorem addresses the convergence of heterogeneous learning dynamics for ZSSGs.

\begin{theorem}\label{thm:SG}
Consider a ZSSG characterized by $\mathcal{M} = \langle S,(A^i,r^i)_{i=1}^2,p,\gamma\rangle$. Suppose that agents follow a learning dynamic from Algorithm Family \ref{tab:algoSG} and the local step sizes $\alpha_k^i$ and $\beta_k^i$'s satisfy, resp., Assumptions \ref{assume:step}, \ref{assume:stepbeta} and \ref{assume:stepbetacompare}. Furthermore, 
\begin{itemize}
\item the local step sizes for the local Q-function update are comparable such that $\lim_{k\rightarrow\infty}\alpha_k^i/\alpha_k^j\in (0,\infty)$ and, without loss of generality, we assume that $\alpha_k^1/\alpha_k^2\rightarrow d\in (0,1]$ as $k\rightarrow\infty$,
\item the local step sizes for the value function updates may or may not be comparable.
\end{itemize}
Let $\pi_{t}^i(s) \in \Delta^i$ denote a weighted empirical average of the agent $i$'s actions at state $s$ until the $t$th visit. Let $a_{t}(s) \in A^i$ denote the action played by agent $i$ at the $t$th visit to state $s$. Then, $\pi_{t}(s)$ evolves according to
\be\label{eq:beliefSG}
\pi_{t+1}^i(s) = \pi_{t}^i(s) + \alpha_{t}^j(a^i_{t}(s) - \pi_{t}^i(s))
\ee
with arbitrary $\pi_0^i(s)$. Let $Q_t^i(s,a)\in\mathbb{R}$ denote the global Q-function associated with the value function estimates and defined by
\be\label{eq:Qt}
Q_t^i(s,a) := r^i(s,a) + \gamma \sum_{s_+\in S} p(s_+\mid s,a) v_{t}^i(s_+).
\ee
If Assumption \ref{assume:state} also holds and the discount factor $\gamma \in [0,d/2)$, then we have $\|q_t^i(s,\cdot) - Q_t^i(s,\cdot)\pi_t^j(s)\|_2 \rightarrow 0$ as $t\rightarrow \infty$ and
\be\nn
\limsup_{t\rightarrow\infty}|v_t^i(s)-v_*^i(s)| \leq \frac{2d+2\gamma - 3\gamma d}{(1-\gamma)(d-2\gamma)}\sum_{l=1}^2\tau^l\log|A^l|
\ee
almost surely for each $i=1,2$ and $s\in S$.
\end{theorem}

A sketch of the proof is as follows: We first follow similar steps with the proof of Theorem \ref{thm:G} to characterize the limit set of the local Q-function estimates and the empirical averages of the actions as if the value function estimates are stationary due to the two-timescale learning assumption. However, this characterization is not enough to show that the update of the value function estimates have contraction-like property. Therefore, we first focus on showing the asymptotic non-negativity of $v_t^i(s)-v_*^i(s)$ as $t\rightarrow \infty$ and then use this result to establish the contraction property to complete the proof.

\begin{myproof}
The local Q-function estimates again remain bounded in Algorithm Family \ref{tab:algoSG} since the step sizes in steps \ref{step:qupdateSG} and \ref{step:vupdateSG} take values in $[0,1]$. Due to \eqref{eq:lowerbound}, Assumption \ref{assume:state} and the boundedness of the estimates yield that any state gets visited infinitely often, i.e., $t_k^s\rightarrow \infty$ almost surely for each $s$ as $k\rightarrow\infty$. Furthermore, the decaying step sizes imply that the normalization at step \ref{step:normalization} becomes redundant eventually for each $s$. Hence, for any $\epsilon>0$, there exists a (path-dependent) stage $K_{\epsilon}$ such that the normalization is redundant, and the iterates $q_{t_k^s}^i(s,\cdot)\in \Xi_q^i$ and $v_{t_k^s}^i(s) \in \Xi_v^i$, where $\Xi_q^i$ and $\Xi_v^i$ are compact subsets of, resp., $\mathbb{R}^{|A^i|}$ and $\mathbb{R}$, for all $k\geq K_{\epsilon}$. 

Define $\bar{v}_k^i(s) = v_{t_k^s}^i(s)$ for all $s$.
Following the proof in \citep{ref:Sayin21}, we fix state $s$ and let $k_t$ denote the stage at which state $s$ gets visited for the $t$th times. The changes between the $t$th and the $(t+1)$st visits at $K_{\epsilon}\leq k_t < k_{t+1}$ can be written as
\begin{align}
&\begin{bmatrix}
q_{t+1}^i(s,\cdot) \\ \pi_{t+1}^j(s) \\ \bar{v}_{k_{t+1}}^i(\cdot)
\end{bmatrix} = (1-\alpha_{t}^i)\begin{bmatrix}
q_{t}^i(s,\cdot) \\ \pi_{t}^j(s) \\ \bar{v}_{k_t}^i(\cdot)
\end{bmatrix}+  \alpha_{t}^i \left(\begin{bmatrix}
Q^i_{t}(s,\cdot)\mu_{t}^j(s,\cdot) \\ \mu_{t}^j(s,\cdot) \\ \mathbf{0}
\end{bmatrix} + \begin{bmatrix}\uomega_{t}^i(s,\cdot) \\ \oomega_{t}^j(s,\cdot) \\ \mathbf{0} \end{bmatrix}+ \begin{bmatrix} \mathbf{0}\\\mathbf{0} \\ \varepsilon_{t}^i(s,\cdot)\end{bmatrix}\right),\nn
\end{align}
where the response $\mu_{t}^j(s,\cdot)$ is either the best or smoothed best response to $q_{t}^j(s,\cdot)$ depending on agent $j$'s preference. In the update of $\pi_{t}^j(s)$, the noise term $\oomega_{t}^j(s,\cdot) = a_t^j(s) - \mu_{t}^j(s,\cdot)$. In the update of $q_{t}^i(s,\cdot)$, the noise term $\uomega_{t}^i(s,\cdot) = Q_{t}^i(s,\cdot)a_t^j(s) - Q_{t}^i(s,\cdot)\mu_{t}^j(s,\cdot)$ if agent $i$ observes the opponent action and knows the model. If the agent does not observe the opponent action or know the model, then we have
\begin{align}
\uomega_{t}^i(s,a^i) =&\; - \big(\Exp_{a^j \sim \mu_{t}^j(s,\cdot)}[Q_{t}^i(s,a^i,a^j)] - q_{t}^i(s,a^i)\big)\nn\\
&+\frac{\mathbb{I}_{\{a^i=a_{t}^i(s)\}}}{\mu_{t}^i(s,a^i)}(r^i(s,a^i,a_{t}^j(s)) + \gamma v_{k_t}^i(s_{k_t+1}) - q_{t}^i(s,a^i)).\nn
\end{align}
Different from \eqref{eq:uw}, here $\uomega_{t}^i(s,\cdot)$ depends also on $s_{k_{t}+1}\sim p(\cdot\mid s_{k_t},a_{t}^1(s),a_t^2(s))$.
Lastly, the error term $\varepsilon_t^i(s,\cdot):S\rightarrow \mathbb{R}$ is defined by
\be\label{eq:err}
\varepsilon_{t}^i(s,s') = \frac{\bar{v}_{k_{t+1}}^i(s') - \bar{v}_{k_t}^i(s')}{\alpha_{t}^i}.
\ee
The numerator depends on the random $k_{t+1}$ and the denominator $\alpha_t^i$ goes to zero as $t\rightarrow \infty$. 

Note that Assumptions \ref{assume:stepbetacompare} and \ref{assume:state} are equivalent to, resp., \citep[Assumptions 1-ii and 2-i]{ref:Sayin21}. Therefore, we can invoke \citep[Lemma 2]{ref:Sayin21} to conclude that $\{\varepsilon_{t}^i\}_{t\geq 0}$ is asymptotically negligible almost surely. 

Let $y_t = [q_{t}^i(s,\cdot);\pi_{t}^j(s);\bar{v}_{k_t}^i(\cdot)]_{i=1}^2\in Y$, which depends on state $s$ implicitly and $Y= \prod_{i=1}^2 \Xi_q^i \times \Delta^j \times \Xi_v^i$ is a compact subset of $\mathbb{R}^{2(m+|S|)}$. The parameter $y_t$ satisfies
\be
y_{t+1} - y_t - \alpha_t (\omega_t + \bar{e}_t) \in \alpha_t \cdot \bar{F}(y_t).
\ee
where 
\be
\omega_t = \left[\uomega_t^1(s,\cdot);\oomega_t^2(s,\cdot);\mathbf{0};\frac{\alpha_t^2}{\alpha_t^1}\uomega_t^2(s,\cdot);\frac{\alpha_t^2}{\alpha_t^1}\oomega_t^1(s,\cdot);\mathbf{0}\right],
\ee 
the error term
\be
\bar{e}_t = \begin{bmatrix} \mathbf{0} \\ \varepsilon_t^1(s,\cdot) \\ \left(\frac{\alpha_t^2}{\alpha_t^1} - \frac{1}{d}\right)\left(\begin{bmatrix} Q_{t}^2(s,\cdot)\mu_{t}^1(s) \\ \mu_{t}^1(s)\end{bmatrix} - \begin{bmatrix} q_{t}^2(s,\cdot) \\ \pi_{t}^1(s)\end{bmatrix}\right)\\ \frac{\alpha_t^2}{\alpha_t^1}\varepsilon_t^2(s,\cdot)\end{bmatrix},
\ee 
and $\bar{F}(\cdot)$ is defined by
\be
\bar{F}(y) = \left\{\begin{bmatrix} f^{12} \\ \mathbf{0} \\ \frac{1}{d} \cdot f^{21} \\ \mathbf{0} \end{bmatrix}: f^{ij} \in \bar{F}^{ij}(y)\mbox{ for }i=1,2\right\}
\ee
and $\bar{F}^{ij}(y)$ is as described in \eqref{eq:Fijbest} (or \eqref{eq:Fijsmooth}) with $R^i$ is replaced with $Q^i(\cdot) = r^i(s,\cdot) + \gamma \sum_{s_+}p(s_+|s,\cdot) v^i(s_+)$ if agent $j$ plays according to the best (or the smoothed best) response. Since $F^{ij}$ is a Marchaud map, $\bar{F}^{ij}$ and $\bar{F}$ are Marchaud maps.

As in the proof of Theorem \ref{thm:G}, the noise terms satisfy $\Exp\big[[\oomega_{t}^i;\uomega_{t}^i] | \mathcal{F}_{k_t}\big] = \mathbf{0}$ and $\Exp\big[\|[\oomega_{t}^i;\uomega_{t}^i]\|^2|\mathcal{F}_{k_t}\big] < W$ for some finite $W$ for all $t$ such that $k_t\geq K_{\epsilon}$, where $\mathcal{F}_{k_t} := \sigma(q_l^i(s,\cdot),\pi_l^i(s),v_{k_l}^i(\cdot)\mbox{ for } i=1,2, \mbox{ and }l \leq t)$. Since $\{\varepsilon_{t}^i\}_{t\geq 0}$ is asymptotically negligible almost surely, $\lim_{k\rightarrow\infty}\alpha_k^1/\alpha_k^2 =d\in(0,1]$ yield that $\bar{e}_t$ is also asymptotically negligible almost surely. Therefore, the iterate $y_t$ converges to a compact connected internally chain transitive set of the differential inclusion $\dot{y}\in \bar{F}(y)$.

Let $y:\mathbb{R}\rightarrow Y$ be the solution to $\dot{y}\in \bar{F}(y)$ with initial point $y_o \in Y$ such that $y(0) = y_o$ and $\frac{dy(t)}{dt} \in \bar{F}(y(t))$ for almost every $t\in \mathbb{R}$. Define $q^i:\mathbb{R}\rightarrow \Xi_q^i$, $\pi^j:\mathbb{R}\rightarrow \Delta^j$, and $v^i(s):\mathbb{R}\rightarrow \Xi_v^i$ such that $y(t) = [q^i(t);\pi^j(t);[\bar{v}^i(\tilde{s})(t)]_{\tilde{s}\in S}]_{i=1}^2$. Then, for almost every $t\in \mathbb{R}$, we have
\begin{subequations}\label{eq:DISG}
\begin{align}
&\frac{dq^i(t)}{dt} = d^i(Q^i(t) \mu^j(t) - q^i(t))\\
&\frac{d\pi^j(t)}{dt} = d^i(\mu^j(t) - \pi^j(t))\\
&\frac{d\bar{v}^i(\tilde{s})(t)}{dt} = 0\quad \forall \tilde{s}\in S\label{eq:vdot}
\end{align}
\end{subequations}
for each $i=1,2$ and $j\neq i$, where $\mu^j(t) \in \BR^j(q^j(t))$ (and $\mu^j(t) = \sBR^j(q^j(t)))$ if agent $j$ plays according to the best response (and the smoothed best response), and $d^1 = 1$ and $d^2 = 1/d \in [1,\infty)$. The definition of $Q^i$ and \eqref{eq:vdot} yield that there exists $\bar{Q}^i = Q^i(t)$ for all $t$. Therefore, \eqref{eq:DISG} reduces to \eqref{eq:DIG} and we can use the Lyapunov function \eqref{eq:V}, formulated in the proof of Theorem \ref{thm:G}. To this end, we define 
\begin{align}
\bdelta_t^i(s) := &\;\mu_{t}^i(s,\cdot)^Tq_{t}^i(s,\cdot) + \tau^i H^i(\mu_{t}^i(s,\cdot)) - \tau^j H^j(\pi_{t}^j(s)) - \val^i(Q_{t}^i(s,\cdot)),\label{eq:bdelta}\\
\bc_t(s) := &\;\lambda \|\bQ_t(s,\cdot)\|_{\max} - \sum_{i=1}^2\val^i(Q_{t}^i(s,\cdot))\label{eq:bc}
\end{align}
for some $\lambda > 1$, where $\bQ_t(s,\cdot) \equiv Q_{t}^1(s,\cdot)+Q_{t}^2(s,\cdot)^T$ for each $s$. Then, the Lyapunov function yields that
\begin{subequations}
\begin{align}
&\bdelta_t^i(s) \geq \|q_{t}^i(s,\cdot)-Q_{t}^i(s,\cdot)\pi_{t}^j(s)\|_2\rightarrow0\label{eq:qlower}\\
&\bdelta_{t}^1(s) + d \cdot\bdelta_{t}^2(s) \leq \bc_t(s) + \ane,\label{eq:zoomin}
\end{align}
\end{subequations}
for some asymptotically negligible error $(\ane)$ decaying to zero almost surely. 

We can also write \eqref{eq:zoomin} as
\be\label{eq:zoomin2}
\sum_{i=1}^2\bdelta_t^i(s) \leq \bc_t(s) + (1-d) \bdelta_t^2(s) + \ane.
\ee
On the right-hand side, we have $\bdelta_t^2(s)$. 
By \eqref{eq:qlower}, Lemma \ref{lem:contraction} yields that we can bound $\bdelta_t^2(s)$ from above by
\begin{align}\label{eq:delta2}
\bdelta_t^2&(s)\leq  \sum_{i=1}^2\bdelta_t^i(s)  - \|q_t^1(s,\cdot) - Q_t^1(s,\cdot)\pi_t^2(s)\|_2+ \val^1(Q_t^1(s,\cdot))+ \val^2(Q_t^2(s,\cdot)) - \min_a\bQ_t(s,a).
\end{align}
Therefore, by \eqref{eq:bc}, \eqref{eq:qlower}, \eqref{eq:zoomin2} and \eqref{eq:delta2}, we obtain
\begin{align}\label{eq:qsum}
\sum_{i=1}^2\bdelta_t^i(s) \leq &\;\frac{\lambda}{d}\|\bQ_t(s,\cdot)\|_{\max} - \frac{1-d}{d}\min_{a}\bQ_t(s,a) - \val^1(Q_t^1(s,\cdot))- \val^2(Q_t^2(s,\cdot)) + \ane.
\end{align}

By \eqref{eq:entropy}, \eqref{eq:bdelta} and \eqref{eq:qsum}, we have
\begin{align}\label{eq:muqsum}
\sum_{i=1}^2 \mu_t^i(s,\cdot)^Tq_t^i(s,\cdot) \leq &\;\frac{\lambda}{d}\|\bQ_t(s,\cdot)\|_{\max} - \frac{1-d}{d}\min_{a}\bQ_t(s,a)+\xi + \ane,
\end{align}
where $\xi:=\sum_{i=1}^2 \tau^i\log|A^i|$.
We highlight the role of $\mu_t^i(s,\cdot)^Tq_t^i(s,\cdot)$ in step \ref{step:vupdateSG} in Algorithm Family \ref{tab:algoSG}. Here, the second term $\min_a\bQ_t(s,a)$ on the right-hand side poses a challenge to establish the contraction property. By \eqref{eq:Qt}, the fact that $r^1(s,a)+r^2(s,a) = 0$ for all $(s,a)$ implies that
\begin{align}
\bQ_t(s,a) &= \gamma\sum_{s_+\in S} p(s_+\mid s,a) \sum_{i=1}^2 v_t^i(s_+)
\label{eq:Qtsum}
\end{align}
for all $(s,a)$. Correspondingly, we focus on the value function update to address $\min_a \bQ_t(s,a)$.

Let $t$ denote $t_{k-1}^s$ with implicit dependence on state and stage for notational simplicity, and define $\tv_{k-1}^i(s) := v_{t}^i(s) - v_*^i(s)$ for all $k$.  Then, $\tilde{v}_k^i(s)$ remains unchanged if $s$ does not get visited and evolves according to
\begin{align}\nn
\tv_{k}^i(s) = &\;(1-\bar{\beta}_{k-1}^i(s))\tv_{k-1}^i(s)+ \bar{\beta}_{k-1}^i(s)\left(\mu_t^i(s)^Tq_{t}^i(s,\cdot) - v_*^i(s)\right),\label{eq:vtilde}
\end{align}
where the step size $\bar{\beta}_k^i(s) := \mathbb{I}_{\{s=s_k\}}\beta_{t}^i\in[0,1]$. Note that 
\begin{align}
\mu_t^i(s)^Tq_{t}^i(s,\cdot) - v_*^i(s) \geq&\;\val^i(Q_t^i(s,\cdot))-\maxmin^i(Q_t^i(s,\cdot)) + \maxmin^i(Q_t^i(s,\cdot)) - \maxmin^i(Q_*^i(s,\cdot))\nn\\
&+\bdelta_t^i(s) -\tau^iH^i(\mu_t^i(s,\cdot)) + \tau^jH^j(\pi_t^j(s))
\end{align} 
which follows from \eqref{eq:bdelta} and $v_*^i(s)= \maxmin^i(Q_*^i(s,\cdot))$. Based on \eqref{eq:entropy}, \eqref{eq:Qstar}, \eqref{eq:Qt} and \eqref{eq:qlower}, Lemma \ref{lem:contraction} yields that
\begin{align}\label{eq:mutlower}
\mu_t^i(s)^Tq_{t}^i(s,\cdot)  - &\;v_*^i(s) \geq \gamma \min_{s_+\in S}\{\tilde{v}_k^i(s_+)\}-\ue_k^i(s)
\end{align}
for some $\lim_{k\rightarrow\infty}\ue_k^i(s) \xi$ almost surely. Hence, $\tv_k^i(s)$'s satisfy
\begin{align}\nn
\tv_{k}^i(s) \geq &\;(1-\bar{\beta}_{k-1}^i(s))\tv_{k-1}^i(s)+ \bar{\beta}_{k-1}^i(s)\left(\gamma \min_{s_+\in S}\{\tilde{v}_k^i(s_+)\}-\ue_k^i(s)\right).\label{eq:rec}
\end{align}

Since $t_k^s\rightarrow\infty$ as $k\rightarrow\infty$ almost surely and $\beta_k^i$ satisfies Assumption \ref{assume:stepbeta}, we can invoke Lemma \ref{lem:async} (from Appendix \ref{app:prelim}) for \eqref{eq:rec} to conclude that
\be\label{eq:tvliminf}
\liminf_{k\rightarrow\infty} \tv_{k}^i(s) \geq - \frac{\xi}{1-\gamma}\quad\forall s
\ee
almost surely.
Since
$v_*^1(s)+v_*^2(s) = 0$ for all $s$, we have $\tv_k^1(s)+\tv_k^2(s) = v_{t_k^s}^1(s)+v_{t_k^s}^2(s)$. By \eqref{eq:Qtsum} and \eqref{eq:tvliminf}, we obtain
\be\label{eq:Qsumliminf}
\liminf_{t\rightarrow \infty}\min_{a} \bQ_t(s,a) \geq - \frac{2\gamma\xi}{1-\gamma}.
\ee
Then, by \eqref{eq:muqsum} and \eqref{eq:Qsumliminf}, we have
\begin{align}
&\sum_{i=1}^2 (\mu_t^i(s,\cdot)^Tq_t^i(s,\cdot) - v_*^i(s))\leq \frac{\lambda}{d}\|\bQ_t(s,\cdot)\|_{\max} + \left(1 + \frac{1-d}{d}\frac{2\gamma}{1-\gamma}\right)\xi + \ane.
\end{align}
Furthermore, the lower bound \eqref{eq:mutlower} and \eqref{eq:tvliminf} yield that
\be\label{eq:b1}
\mu_t^j(s,\cdot)^Tq_t^j(s,\cdot) - v_*^j(s) \geq - \frac{\xi}{1-\gamma} + \ane.
\ee
Therefore, we obtain
\begin{align}\label{eq:b2}
&\mu_t^i(s,\cdot)^Tq_t^i(s,\cdot) - v_*^i(s)\leq \frac{\lambda}{d}\|\bQ_t(s,\cdot)\|_{\max} + \left(1 + \frac{1}{1-\gamma}+\frac{1-d}{d}\frac{2\gamma}{1-\gamma}\right)\xi + \ane
\end{align}
Since $v_*^1(s)+v_*^2(s) = 0$, \eqref{eq:Qtsum} yields that
\begin{align}\label{eq:b3}
\bQ_t(s,a) &= \gamma\sum_{s_+\in S} p(s_+\mid s,a) \sum_{i=1}^2 (v_t^i(s_+)- v_*^i(s_+))\nn\\
&\leq \gamma \max_{s_+\in S} \sum_{i=1}^2|v_t^i(s_+)- v_*^i(s_+)|\nn\\
&\leq 2\gamma \max_{s_+\in S} \max_{i=1,2}|v_t^i(s_+)- v_*^i(s_+)|
\end{align}
for any $(s,a)$. Combining \eqref{eq:b1}, \eqref{eq:b2} and \eqref{eq:b3}, we obtain
\begin{align}\label{eq:gammalambda}
|\mu_t^i(s,\cdot)^Tq_t^i(s,\cdot) - v_*^i(s)|\leq&\; \frac{2\gamma\lambda}{d} \max_{s_+\in S} \max_{i=1,2}|v_t^i(s_+)- v_*^i(s_+)| \nn\\
&+ \left(1 + \frac{1}{1-\gamma}+\frac{1-d}{d}\frac{2\gamma}{1-\gamma}\right)\xi + \ane
\end{align}
for each $s$ and $t$. Define $y_k = [\tilde{v}_k^1(\cdot);\tilde{v}_k^2(\cdot)]\in\mathbb{R}^{2|S|}$ such that $\|y_k\|_{\infty} = \max_{s_+\in S} \max_{i=1,2}|v_{t_k^s}^i(s_+)- v_*^i(s_+)|$. By \eqref{eq:vtilde} and \eqref{eq:gammalambda}, $y_k$ satisfies the inequalities in Lemma \ref{lem:asynctwosided} provided in Appendix \ref{app:prelim}. Recall that $t_k^s\rightarrow\infty$ as $k\rightarrow\infty$ almost surely and $\beta_k^i$ satisfies Assumption \ref{assume:stepbeta}. If $2\gamma\lambda/d < 1$, then we can invoke Lemma \ref{lem:asynctwosided} to conclude that
\begin{align}
\limsup_{k\rightarrow\infty} |\tilde{v}_k^i(s)| &\leq \frac{1}{1-2\gamma\lambda/d}\left(1 + \frac{1}{1-\gamma}+\frac{1-d}{d}\frac{2\gamma}{1-\gamma}\right)\xi\nn\\
&= \frac{2d+2\gamma-3\gamma d}{(1-\gamma)(d-2\lambda \gamma)}\xi
\end{align}
almost surely for each $i=1,2$ and $s\in S$.
This completes the proof.
\end{myproof}

The following corollary to Theorem \ref{thm:SG} shows the rationality of the dynamics from Algorithm Family \ref{tab:algoSG}.

\begin{corollary}\label{cor:rationalSG}
Consider a ZSSG characterized by $\mathcal{M} = \langle S,(A^i,r^i)_{i=1}^2,p,\gamma\rangle$. Suppose that agent $i$ follow a learning dynamic from Algorithm Family \ref{tab:algoSG} while the opponent $j$ plays according to fully mixed strategy $\pi^j:S\rightarrow \mathrm{int}(\Delta^j)$. The local step sizes $\alpha_k^i$ and $\beta_k^i$'s satisfy, resp., Assumptions \ref{assume:step} and \ref{assume:stepbeta}. Furthermore, we have $\beta_k^i/\alpha_k^i\rightarrow 0$ as $k\rightarrow\infty$. 

Let $\pi_{t}^i(s) \in \Delta^i$ denote a weighted empirical average of the agent $i$'s actions at state $s$ until the $t$th visit. Let $a_{t}(s) \in A^i$ denote the action played by agent $i$ at the $t$th visit to state $s$. Then, $\pi_{t}(s)$ evolves according to \eqref{eq:beliefSG} with arbitrary initializations. 
If Assumption \ref{assume:state} also holds, then $q_t^i(s,a^i)$ tracks $$\Exp \left[r^i(s,a^i,a^j) + \gamma \sum_{s_+} p(s_+|s,a^i,a^j) v_{t}^i(s_+)\right],$$ where the expectation taken with respect to $a^j\sim\pi^j(s)$ and 
\be\nn
\limsup_{t\rightarrow\infty}|v_t^i(s)-v_*^i(s)| \leq \frac{(2-\gamma)\tau^i\log|A^i|}{(1-\gamma)(1-2\gamma)}
\ee
almost surely for each $s\in S$.
\end{corollary}

\begin{myproof}
As in the proof of Corollary \ref{cor:rationalG}, we can consider that agent $i$ is playing a ZSSG against an opponent (other than agent $j$) following the same dynamic with singleton action spaces at each state and view agent $j$ as the Nature. Then, the proof follows from Theorem \ref{thm:SG}. 
\end{myproof}

\begin{remark}\label{remark:loose}
We address different step sizes $\beta_k^i\neq \beta_k^j$ via the upper bounds in \eqref{eq:b3}. Correspondingly, we can attain tighter error bounds for the cases where $\beta_k^i=\beta_k^j$ in Theorem \ref{thm:SG} and Corollary \ref{cor:rationalSG}, as in \citep{ref:Sayin20,ref:Sayin21}. 
\end{remark}

\begin{figure}[t!]
  \centering
  \begin{subfigure}{.45\columnwidth}
  \includegraphics[width=\columnwidth]{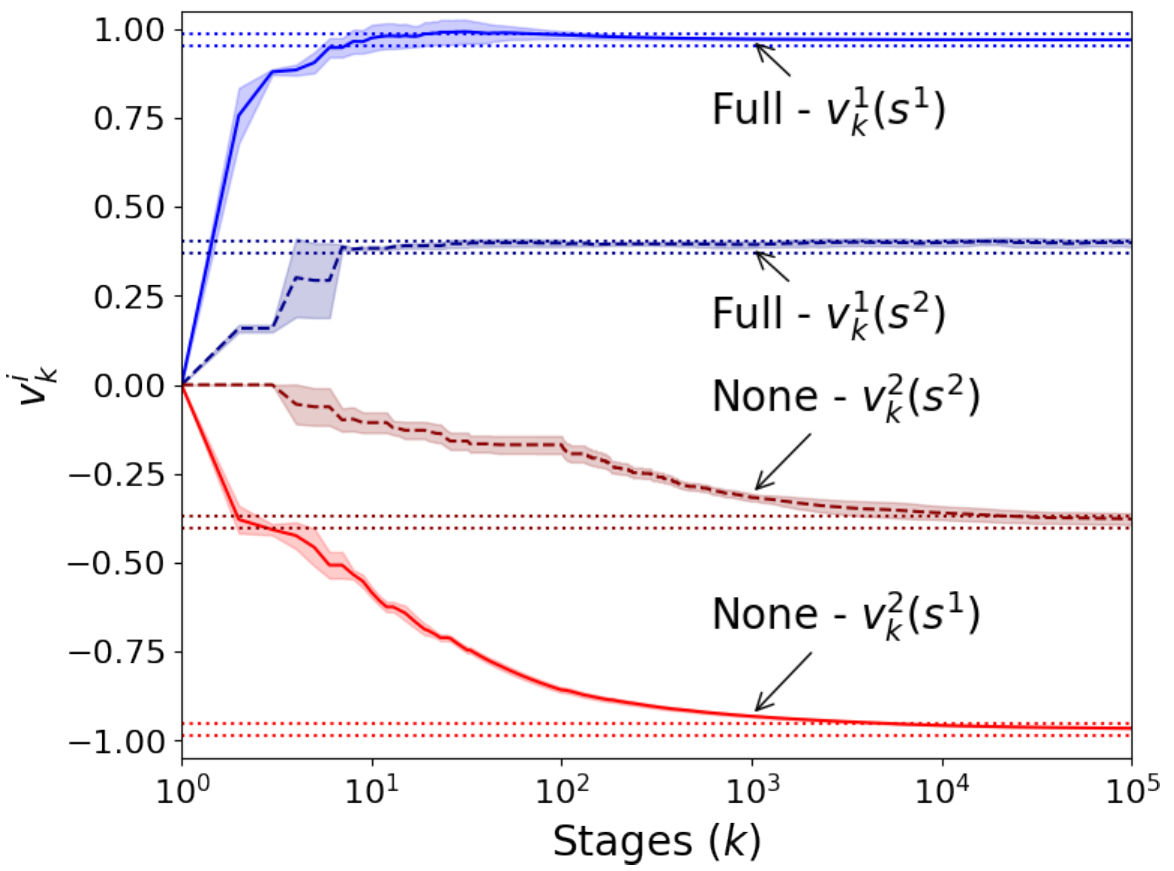}%
  \caption{Full Access vs No Access}%
  \label{subfiga}%
  \end{subfigure}
  \hspace{.15in}
  \begin{subfigure}{.45\columnwidth}
  \includegraphics[width=\columnwidth]{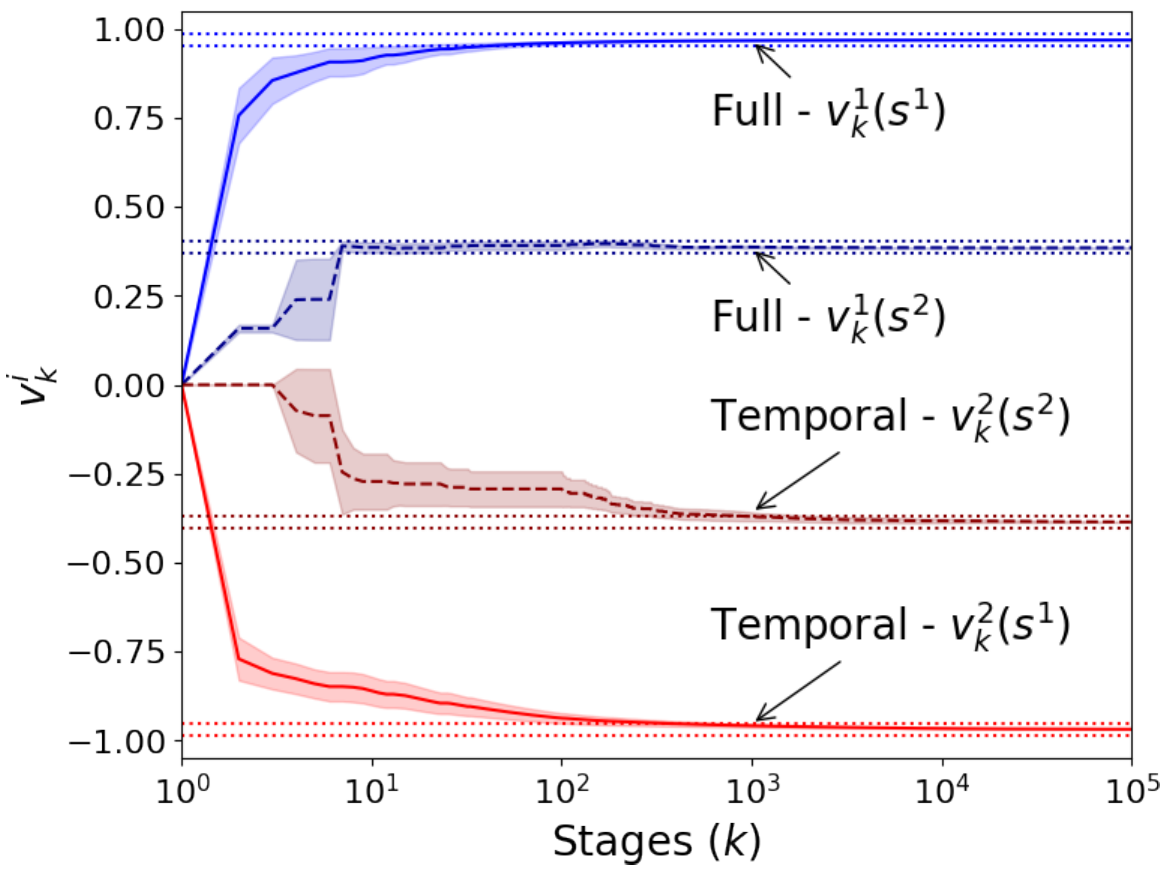}%
  \caption{Full Access vs Temporal Access}%
  \label{subfigc}%
  \end{subfigure}\\
   \vspace{.15in}
    \begin{subfigure}{.45\columnwidth}
  \includegraphics[width=\columnwidth]{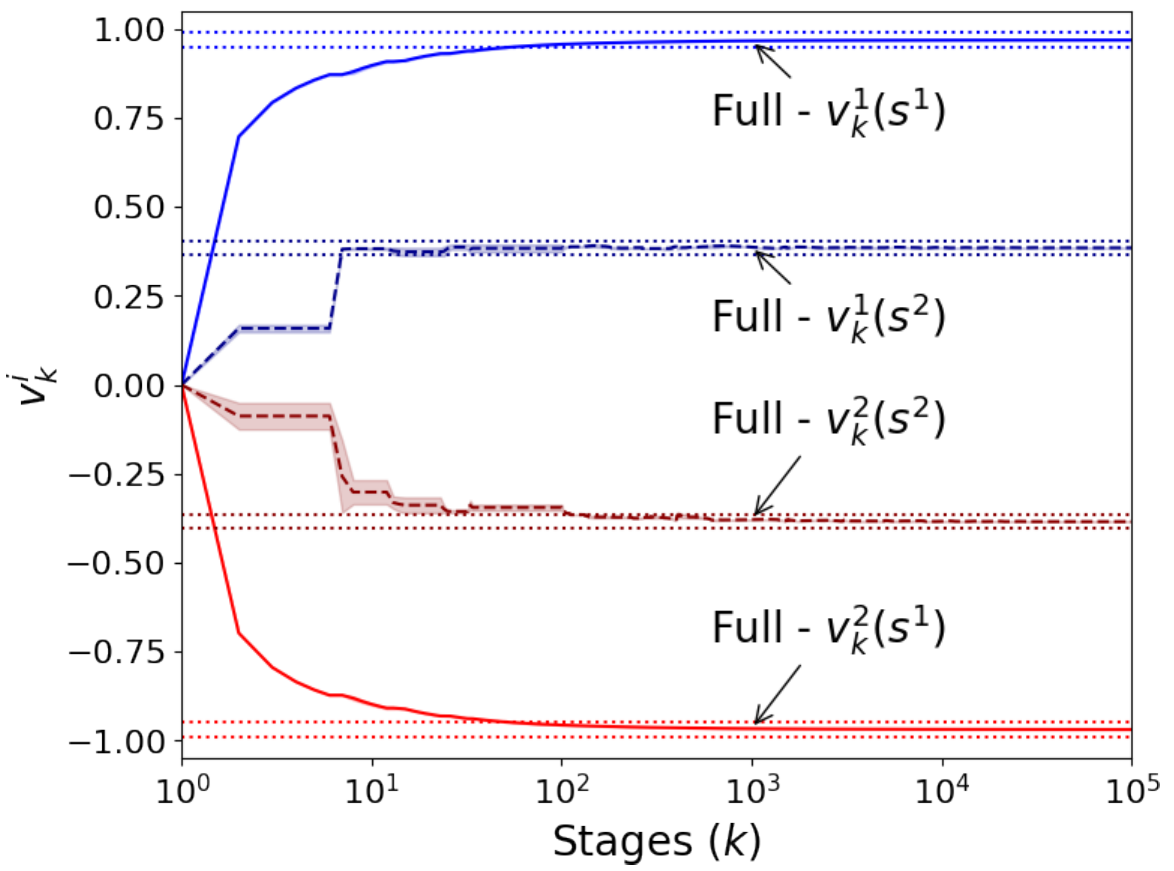}%
  \caption{Different Learning Rates}%
  \label{subfigb}%
  \end{subfigure}
  \caption{The evaluation of estimated value functions in a ZSSG with two states and two actions, where $S = \{s^1, s^2\}$, $\gamma = 0.3$ and $\tau = 0.002$. Each sub-figure corresponds to Scenarios $1$, $2$, and $3$, respectively.   ``Full",  ``Temporal" and ``None" refer to whether the agent has full, temporal (with $\theta^i=0.5$) or no access to the opponent's actions. The different colors and line styles represent different agents and states. The shaded areas represent the standard deviation of the value function estimates. The dotted lines represent the bounds on the value function estimates according to Theorem \ref{thm:SG}.}
  \label{fig:simulations}
  \end{figure}

\section{Illustrative Examples} \label{sec:examples}

As a proof of concept, we consider a ZSSG with two states $|S|=2$ and two actions $|A^i|=2$ for $i=1,2$. We examine three distinct scenarios covering the heterogeneities in terms of step sizes, exploration, access to opponent actions, and model knowledge. In all of the scenarios, reward functions and transition kernel are the same for comparison across scenarios. The rewards are chosen uniformly from $[0, 1]$ and $[0, 0.2]$ for first and second states, respectively, to generate aesthetic plots. The transition probabilities are chosen randomly from a symmetric Dirichlet distribution $\mathrm{Dir}(\alpha)$ with parameters $\alpha = \mathbf{1}$ such that $p(s'|s,a) > 0$ for all $(s,a,s')$. The discount factor is $\gamma=0.3$ and the temperature parameter is $\tau=0.002$ for all scenarios. 

We conducted the simulations on a desktop computer equipped with an Intel Xeon W7-3455 CPU and 128GB RAM. We used Python 3.11.5 and the version 0.0.9 of \texttt{coaction}\footnote{\texttt{coaction} library can be accessed in \url{https://github.com/yigityalin/coaction}}. 
For each scenario, we run $30$ independent trials of the algorithms.

In Figure \ref{fig:simulations}, we present the average behavior of value function estimates for the following scenarios. 

\begin{itemize}
\item \textit{Scenario 1. Full Access vs No Access:}
In this scenario, agent $1$ has access to agent $2$'s actions and plays according to smoothed best response, while agent $2$ does not observe agent $1$'s actions and instead plays according to the rewards received. This scenario can also be interpreted as a comparison of belief-based versus payoff-based learning dynamics. A special case of this interaction can be seen in the comparison between smoothed fictitious play and individual Q-learning. 
\item \textit{Scenario 2. Full Access vs Temporal Access:}
In this setup, agent $1$ has full information access while agent $2$ has access to the opponent's actions with probability $\theta^2=0.5$. Comparing with Scenario 1, we observe that temporal access to the opponent's actions lead to faster convergence than the no access one. 
\item \textit{Scenario 3. Different Learning Rates:}
In this setup, both agents have access to the opponent's action and play according to smoothed best response while the heterogeneity stems from the different learning rates such that they update their iterates with step sizes $\alpha^{1}_k = 1/k^{0.96}$, $\alpha^{2}_k = 1/(0.92k)^{0.96}$ and $\beta^{1}_k = 1/k$, $\beta^{2}_k = 1/(0.96k)$ satisfying standard assumptions listed in Theorem \ref{thm:SG}. 
\end{itemize}

In all three scenarios, the agents' value function estimates are contained within the bounds indicated by Theorem \ref{thm:G}. Since the size of the game, i.e., $|S|$ and $|A^i|$, is small, we observe small variance in the value function estimates across independent trials, especially in Scenario 3 where both agents know the model and have full access to the opponent's actions.

\section{Conclusion}\label{sec:conclusion}
We studied heterogeneous learning dynamics for the repeated play of two-agent (near) zero-sum games and two-agent zero-sum stochastic games. We presented two families of algorithms addressing heterogeneities in terms of learning rates, exploration vs exploitation, (temporal) access to opponent actions (i.e., belief-based vs payoff-based) and model-based vs model-free. Any dynamic from these families are uncoupled and rational. We also showed that any mixture of dynamics from these families  converge to equilibrium (or near equilibrium if there is experimentation) in zero-sum (stochastic) games under standard assumptions on step sizes and the transition kernel, additionally sufficiently small discount factor if the agents use different step sizes. 

Possible future research directions include addressing heterogeneous learning $i)$ for non-zero-sum (stochastic) games, $ii)$ for actor-critic-type or gradient-based learning dynamics rather than best-response ones, and $iii)$ for transient regimes rather than steady state one via non-asymptotic analysis to identify the impact of heterogeneities on the convergence rates.

\section*{Acknowledgment}
This work was supported by the The Scientific and Technological Research Council of T\"{u}rkiye (TUBITAK) BIDEB 2232-B International Fellowship for Early Stage Researchers under Grant Number 121C124.

\appendix

\section{Preliminary Information}\label{app:prelim}

In the following, we provide preliminary information on stochastic approximation methods for completeness.

\subsection{Stochastic Differential Inclusion Approximation}\label{app:prelimContinuous}

The following is a refinement of the results in \citep[Proposition 1.3 and Theorem 3.6]{ref:Benaim05} to compact domains. Consider a sequence $\{x_k \in X\}_{k\geq 0}$, where $X$ is a compact subset of $\mathbb{R}^m$, evolving according to
\be\label{eq:SI}
x_{k+1} - x_k - \alpha_k(e_k + \omega_k) \in \alpha_k F(x_k)
\ee
where
\begin{itemize}
\item The step size decay, i.e., $\alpha_k\rightarrow 0$ at a rate such that 
$\sum_{k=0}^{\infty} \alpha_k = \infty$ (and $\sum_{k=0}^{\infty} \alpha_k^2 < \infty$ if we have non-zero noise $\omega_k$). 
\item The set-valued function $F(x)\subset \mathbb{R}^m$ is a Marchaud map, i.e., satisfying the following conditions:
\begin{itemize}
\item $F(\cdot)$ is upper semi-continuous, or equivalently, $\mathrm{Graph}(F) = \{(x,y):x\in X, y\in F(x)\}$ is a closed subset of $X\times \mathbb{R}^m$.
\item For all $x\in X$, $F(x)$ is a non-empty, compact, and convex subset of $X$.
\item There exists a $c > 0$ such that for all $x\in X$, we have $\sup_{y\in F(x)} \|y\|\leq c(1+\|x\|)$. 
\end{itemize} 
\item The error term $\|e_k\|\rightarrow 0$ with probability $1$, as in \citep[Theorem 2.3]{ref:Perkins13}.
\item With respect to the filtration $\mathcal{F}_k = \sigma(x_0,\omega_0,\ldots,\omega_{k-1})$, the noise term $\omega_k$ is $\mathcal{F}_{k+1}$-measurable and satisfies $\Exp[\omega_k\,|\, \mathcal{F}_k] = 0$ and $\Exp[\|\omega_k\|^2 \,|\,  \mathcal{F}_k]< W$ for some constant $W$ for all $k\geq 0$ with probability $1$.
\end{itemize}
Then, $x_k$ almost surely converges to a compact connected internally chain transitive set of the differential inclusion
\be\label{eq:DI}
\dot{x} \in F(x).
\ee 
Hence, we can characterize the limit behavior of \eqref{eq:SI} through \eqref{eq:DI} if there exists a continuous function $V:X\rightarrow \mathbb{R}$, called \textit{Lyapunov function} for $\Lambda \subset X$, such that for any absolutely continuous solution $x(t)$ to \eqref{eq:DI}, we have
\begin{itemize}
\item $V(x(t')) < V(x(t))$ for all $t'> t$ and $x(t)\notin \Lambda$
\item $V(x(t')) \leq V(x(t))$ for all $t'> t$ and $x(t)\in \Lambda$
\end{itemize}
and $V(\Lambda) = \{V(x):X\in \Lambda\} \subset \mathbb{R}$ has empty interior, then every internally chain transitive set of \eqref{eq:DI}, and therefore, the limit set of \eqref{eq:SI} are contained in $\Lambda$ \citep[Proposition 3.7]{ref:Benaim05}.

\subsection{Asynchronous Updates}\label{app:prelimDiscrete}

The following result is adopted from Theorem 5.1 in the extended version of \citep{ref:Sayin20} (built upon \citep[Theorem 1]{ref:Tsitsiklis94}) without the stochastic approximation noise term.\footnote{The extended version of \citep{ref:Sayin20} is available at ArXiv:2010.04223.}

\begin{lemma}\label{lem:asynctwosided}
Consider a sequence of vectors $\{y_k\}_{k\geq 0}$ whose $l$th entry, denoted by $y_k(l)$, satisfies the following lower bound
\begin{align*}
&y_{k+1}(n) \leq (1-\beta_k(n)) y_k(n) + \beta_k(n) (\gamma \|y_k\|_{\infty} - \overline{e}_k(n))\\
&y_{k+1}(n) \geq (1-\beta_k(n)) y_k(n) + \beta_k(n) (-\gamma \|y_k\|_{\infty} - \underline{e}_k(n)), 
\end{align*}
where $\gamma \in (0,1)$, the step size $\beta_k(n)\in[0,1]$ decay to zero sufficiently slowly such that $\sum_{k=0}^{\infty}\beta_k(n)=\infty$, and the errors 
$$
\limsup_{k\rightarrow\infty}|\overline{e}_k(n)| \leq c \quad\mbox{and}\quad\limsup_{k\rightarrow\infty}|\underline{e}_k(n)| \leq c
$$ 
for some $c\geq 0$ almost surely for each $n$. Suppose that $\|y_k\|_{\infty}\geq M$ for some $M\geq 0$ for all $k$ and $n$. Then, we have
$$
\limsup_{k\rightarrow\infty} \|y_k\|_{\infty} \leq \frac{c}{1-\gamma}
$$
almost surely.
\end{lemma}

In Lemma \ref{lem:asynctwosided}, the iterates satisfy upper and lower bounds. The following lemma addresses the case where we only have the lower bound and it is a modification of \citep[Lemma 9]{ref:Sayin22d} to address errors that do not decay to zero.

\begin{lemma}\label{lem:async}
Consider a sequence of vectors $\{y_k\}_{k\geq 0}$ whose $n$th entry, denoted by $y_k(n)$, satisfies the following lower bound
\be\nn
y_{k+1}(n) \geq (1-\beta_k(n)) y_k(n) + \beta_k(n) (\gamma \min_{m} y_k(m) - e_k(n)), 
\ee
where $\gamma \in (0,1)$, the step size $\beta_k(n)\in[0,1]$ decay to zero sufficiently slowly such that $\sum_{k=0}^{\infty}\beta_k(n)=\infty$, and the error $e_k(n)\rightarrow c$ for some $c\geq 0$ as $k\rightarrow\infty$ almost surely for each $n$. Suppose that $y_k(n)\geq -M$ for some $M\geq 0$ for all $k$ and $n$. Then, we have
$$
\liminf_{k\rightarrow\infty} y_k(n) \geq -\frac{c}{1-\gamma}
$$
almost surely for each $n$.
\end{lemma}

\begin{myproof}
Define the sequence $\{M^t<0\}_{t=0}^{\infty}$ over a separate timescale by 
$M^{t+1} = (\gamma + 2\epsilon)M^t$ for all $t$,
and $M^0= M$. We set $\epsilon \in (0,(1-\gamma)/2)$ so that $M^t\rightarrow 0$ monotonically from above as $t\rightarrow \infty$. We claim that for each $t=0,1,\ldots$, there exists $k^t\in \mathbb{N}$ such that 
\be\label{eq:induction}
y_k(n) \geq -M^t - \frac{c}{1-\gamma}\quad\mbox{and}\quad -e_k(n) \geq - \epsilon M^t- c 
\ee
for all $k\geq k^t$ and for each $n$. Since $y_k(n)\geq M^0=M$ for all $k$ and $n$, and $e_k(n)\rightarrow c$, there exists such $k^0$. Suppose that \eqref{eq:induction} holds for some $t$. Specific to $t$, define an auxiliary sequence $\{Y_k^t\}_{k\geq k^t}$ by
\be\nn
Y_{k+1}^t(n) = (1-\beta_k(n))Y_k^t(n) + \beta_k(n)\left(-(\gamma+\epsilon)M^t - \frac{c}{1-\gamma}\right)
\ee
and $Y_{k^t}^t(n) = -M^t-\frac{c}{1-\gamma}$ for each $n$. Note that
\be
\gamma \min_{m} y_k(m) - e_k(n) \geq -(\gamma+\epsilon)M^t - \frac{c}{1-\gamma}
\ee 
for all $k\geq k^t$ as \eqref{eq:induction} holds for $t$. Therefore, we have $Y_k^t(n)\leq y_k(n)$ for all $k\geq k^t$ and $n$. Since $Y_k^t(n)\rightarrow -(\gamma+\epsilon)M^t - \frac{c}{1-\gamma}$ and $-(\gamma+\epsilon)M^t \geq -M^{t+1}$, there exists $k^{t+1}\geq k^t$ such that \eqref{eq:induction} holds for $t+1$. Therefore, by induction, we can conclude that \eqref{eq:induction} holds for every $t\geq 0$. Since $M^t \rightarrow 0$, we have the lower bound.
\end{myproof}

\section{Proof of Lemma \ref{lem:contraction}}\label{app:lemproofs}
The contraction property of $\maxmin^i(\cdot)$ is used, e.g., in \citep{ref:Shapley53}. 
Let $(\mu^1,\mu^2)$  and $(\bmu^1,\bmu^2)$ satisfy
\begin{subequations}
\begin{align}
&\val^1(R^1) := (\mu^1)^TR^1\mu^2+ \tau^1 H^1(\mu^1) - \tau^2 H^2(\mu^2)\\
&\val^2(R^2) := (\bmu^2)^TR^2\bmu^1+ \tau^2 H^2(\bmu^2) - \tau^1 H^1(\bmu^1).
\end{align}
\end{subequations}
Then, we have
\begin{align}
\val^1(R^1) + \val^2(R^2) &\leq (\mu^1)^TR^1\bmu^2+ \tau^1 H^1(\mu^1) - \tau^2 H^2(\bmu^2)+(\bmu^2)^TR^2\mu^1+ \tau^2 H^2(\bmu^2) - \tau^1 H^1(\mu^1)\nn\\
&\leq (\mu^1)^T(R^1+(R^2)^T)\bmu^2
\end{align}
since agent $j$ is the minimizer. Since $\mu^1\in \Delta^1$ and $\bmu^2\in \Delta^2$ are probability distributions, we have $(\mu^1)^T(R^1+(R^2)^T)\bmu^2 \leq \overline{r}$. By following similar lines, we can also show that $\val^1(R^1) + \val^2(R^2)\geq \underline{r}$. 

For the second part, let $(\mu^i,\mu^j)$  and $(\bmu^i,\bmu^j)$ satisfy
\begin{subequations}
\begin{align}
&\val^i(R^i) := (\mu^i)^TR^i\mu^j+ \tau^i H^i(\mu^i) - \tau^j H^j(\mu^j)\\
&\maxmin^i(R^i) := (\bmu^i)^TR^i\bmu^j+ \tau^i H^i(\bmu^i) - \tau^j H^j(\bmu^j).
\end{align}
\end{subequations}
Then, we have
\begin{align}
\val^i(R^i) - \maxmin^i(R^i) &\leq (\mu^i)^TR^i\bmu^j+ \tau^i H^i(\mu^i) - \tau^j H^j(\bmu^j)-(\mu^i)^TR^i\bmu^j\nn\\
&\leq \tau^i\log|A^i|
\end{align}
as $-H^j(\bmu^j)\leq 0$ and $H^i(\mu^i)\leq \log|A^i|$. By following similar lines, we can show that $\val^i(R^i) - \maxmin^i(R^i)\geq -\tau^j\log|A^j|$. This completes the proof.

\end{spacing}

\begin{spacing}{1}
\bibliographystyle{plainnat}
\bibliography{mybib}
\end{spacing}

\end{document}